\documentclass{article}

\usepackage{arxiv}

\usepackage[utf8]{inputenc} 
\usepackage[T1]{fontenc}    
\usepackage{hyperref}       
\usepackage{url}            
\usepackage{booktabs}       
\usepackage{amsfonts}       
\usepackage{nicefrac}       
\usepackage{microtype}      
\usepackage{lipsum}		
\usepackage{graphicx}
\usepackage{natbib}
\usepackage{doi}

\title{Potential of Ka-band Range Rate Post-fit Residuals for High-frequency Mass Change Applications}

\author{ \href{https://orcid.org/0000-0001-5550-962X}{\hspace{1mm}Michal Cuadrat-Grzybowski}\thanks{Corresponding author.} \\
	Space Engineering Department\\
	Delft University of Technology\\
	Kluyverweg 1, 2629 HS Delft (the) Netherlands \\
	\texttt{M.Cuadrat-Grzybowski-1@tudelft.nl} \\
	\And
	\href{https://orcid.org/0000-0001-6824-2733}{\hspace{1mm}Joao G. Teixeira da Encarnacao} \\
	Space Engineering Department\\
	Delft University of Technology\\
	Kluyverweg 1, 2629 HS Delft (the) Netherlands \\
	\texttt{j.g.deteixeiradaencarnacao@tudelft.nl} \\
        \And
	\href{https://orcid.org/0000-0002-2018-7373}{\hspace{1mm}Pieter N. A. M. Visser} \\
	Space Engineering Department\\
	Delft University of Technology\\
	Kluyverweg 1, 2629 HS Delft (the) Netherlands \\
	\texttt{p.n.a.m.visser@tudelft.nl} \\
}

\renewcommand{\shorttitle}{}

\hypersetup{
pdftitle={A template for the arxiv style},
pdfsubject={q-bio.NC, q-bio.QM},
pdfauthor={David S.~Hippocampus, Elias D.~Striatum},
pdfkeywords={First keyword, Second keyword, More},
}

\begin{document}
\maketitle

\begin{abstract}
    We present the first extensive analysis of K/Ka-band ranging post-fit residuals of an official Level-2 product, characterised as Line-of-Sight Gravity Differences (LGD), which exhibit and showcase interesting sub-monthly geophysical signals. These residuals, provided by CSR, were derived from the difference between spherical harmonic coefficient least-squares fits and reduced Level-1B range-rate observations. We classified the geophysical signals into four distinct categories: oceanic, meteorological, hydrological, and solid Earth, focusing primarily on the first three categories in this study. In our examination of oceanic processes, we identified notable mass anomalies in the Argentine basin, specifically within the Zapiola Rise, where persistent remnants of the rotating dipole-like modes are evident in the LGD post-fit residuals. Our analysis extended to the Gulf of Carpentaria and Australia during the 2013 Oswald cyclone, revealing significant LGD residual anomalies that correlate with cyclone tracking and precipitation data. Additionally, we investigated the monsoon seasons in Bangladesh, particularly from June to September 2007, where we observed peaks in sub-monthly variability. These findings were further validated by demonstrating high spatial and temporal correlations between gridded LGD residuals and ITSG-Grace2018 daily solutions. Given that these anomalies are associated with significant mass change phenomena, it is essential to integrate the post-fit residuals into a high-frequency mass change framework, with the purpose of providing enhanced spatial resolution compared to conventional Kalman-filtered methods.
\end{abstract}

\keywords{GRACE, K-Band Ranging, post-fit residuals, Line-of-Sight Gravity Difference, high-frequency, geophysics.}

\section{Introduction}
The Gravity Recovery and Climate Experiment (GRACE) and its follow-on (FO) mission, GRACE-FO, have revolutionised our understanding of Earth's gravity field and mass redistribution. These missions provide monthly solutions of spherical harmonic (SH) and mascon (mass concentrations)-based models, which have been widely utilised in hydrology, glaciology, oceanography, and solid Earth dynamics studies \cite{Schmidt2008-dy, Wahr2015, Wouters2019, Xiong2024_earth}. The primary data products from GRACE are made available by the GRACE Science Data System (SDS), composed of the Center for Space Research (CSR), the Jet Propulsion Laboratory (JPL), and GeoForschungsZentrum Potsdam (GFZ), which generate monthly SH models (also known as Level-2 or L2) through the Least-Squares fitting of reduced Level-1B (L1B) observations, resulting in a set of L2 SH or/and mascon-related coefficients (part of the Level-3 data).

Research efforts aimed at improving the temporal resolution of GRACE(-FO) solutions led to the development of 10-day \cite{Lemoine2020_rl05}, weekly \cite{Flechtner2010}, and even daily \cite{Kurtenbach2012} spherical harmonic (SH) and mascon products \cite{Bonin2020}. Currently, daily solutions produced using Kalman filtering techniques, such as the ITSG-Grace2018 solutions \cite{Kurtenbach2012, Mayer_Gurr_T2018} and other studies \cite{Kvas2019, Ramillien2020}, represent the state-of-the-art in terms of temporal resolution. Even though GRACE daily ground coverage (of 15 revolutions) is not sufficient for global gravity recovery, these daily solutions were stabilised by applying a backward and forward Kalman smoother on daily GRACE data with the addition of de-aliasing sub-monthly hydrological background models \cite{Kvas2019}. \citet{Gouweleeuw2018} were the first to use ITSG-Grace2014 daily solutions to demonstrate that GRACE could observe sub-monthly hydrological signals related to severe flooding events during Bangladesh's monsoon seasons. Furthermore, daily solutions were used to constrain high-frequency hydro-meteorological flux models over continents for the first time \cite{Eicker2020-yu}. However as previously mentioned, other than GRACE(-FO) data, these daily models require the assimilation of additional (hydrological) information for stability. CSR developed GRACE-only daily swaths (limited at $\pm 66$ deg latitude) for high-frequency oceanographic applications, which were able to explain 25\%-75\% sub-monthly altimetry-based observations of mass change \cite{Bonin2020, Schindelegger2021}. Furthermore, CSR is currently developing 5-day (5D) solutions, which showed promising results for flood monitoring and forecasting, with hydrological traces from the 2005 Katrina hurricane identified \cite{Rateb2024, Sun2024}. However, increasing temporal resolution inherently leads to a reduction of spatial resolution, leading to a trade-off between the two. This trade-off sparked interest in alternative approaches that bypassed the SH and mascon frameworks.

One method involves analysing Level-1B data directly, and not the resulting SH or mascon solutions. By working with along-orbit observations, with 5s-sampling, researchers have the potential to directly observe mass variations sensed by GRACE. This approach, when combined with corrections for background models, accelerometer scale and bias factors, and initial state vector inaccuracies, could offer the highest spatial and temporal resolutions. \citet{Ghobadi_Far2019} derived analytical expressions that relate gravimetric observables (such as gravity disturbance potential and gravity gradients) from residual L1B data (i.e. range rate and range acceleration). \citet{Allgeyer2022} have demonstrated the benefit of using time-filtered pre-fit range accelerations when deriving mascon models via Tikhonov regularisation. To effectively eliminate the centrifugal acceleration in L1B range acceleration data, \citet{Ghobadi_Far2018} derived an analytical frequency-based transfer function from range acceleration to Line-of-Sight Gravity (LGD), which better localises mass anomalies on the Earth's surface. Using both range acceleration and LGDs has been particularly valuable in detecting high-frequency mass change events, as highlighted by recent studies \cite{Ghobardi_Far2022, Peidou2022}. Additionally, \citet{Han2021-if} used residual L1B LGDs to estimate rapid changes in water storage due to heavy rainfall and flooding in Australia in March 2021, following extreme drought events in 2019–2020. In the context of solid Earth applications, \citet{Han2010_range_rates} first showed the range rate signature in L1B data related to the 2010 Maule earthquake and later  \citet{Ghobadi_Far2020} demonstrated that residual L1B data in the form of LGDs can also localise the loading effects of tsunamis after major earthquake events such as Sumatra 2004 and Tohoku 2011. Lastly, \citet{Peidou2022} performed an entire spatial and temporal characterisation of GRACE data in terms of range accelerations, which showed that post-fit residuals (obtained after the monthly Least Squares gravity inversion) show sub-monthly geophysical signal. All these studies emphasise the presence of sub-monthly signals within the GRACE L1B data, revealing high-frequency mass change processes by avoiding the SH/mascon monthly representation.

In this study, we consider RL06 CSR L1B post-fit residuals, which are the difference between the pre-fit residuals and the least-squares SH fit, along with additional corrections for "common" and "local” parameters, which is a terminology used by, e.g., \cite{Pini2012}. The pre-fit residuals are the difference between the (preprocessed) KBR measurements and predicted inter-satellite distance generated on the basis of a comprehensive background force model.  As these post-fit residuals represent the errors of a widely used L2 product, the research question that this study answers is 
\begin{center}
    \textit{How can high-frequency geophysical signals within GRACE Level-1B post-fit residuals be identified, characterised, and validated using spatial analysis and daily solutions?}
\end{center}
We characterise and analyse the high-frequency signals in post-fit residuals to gain insights into sub-monthly mass change phenomena.  These residuals reveal important information that the monthly SH coefficients cannot explain, in addition to unmodelled effects from background models and accelerometer calibration parameters \cite{Peidou2022}. The novelty of this study lies in the in-depth spatial and temporal analysis of sub-monthly signals in CSR’s L1B post-fit residuals, offering a more detailed exploration of geophysical phenomena compared to previous studies \cite{Ghobardi_Far2022, Peidou2022}, and the characterisation of error of an official L2 data product. We also show the high resolution with which GRACE can effectively resolve and differentiate various geophysical categories of sub-monthly variations in Earth's time-variable gravity field.

The paper is organised as follows: section \ref{sec:data} describes the data utilised in the study, while section \ref{sec:methods} outlines the methodology employed for analysing high-frequency signals. Section \ref{sec:results} focuses on the analysis of KBR residual L1B data in the context of sub-monthly mass change variations, and section \ref{sec:conclusions} presents the conclusions drawn from this investigation.

\section{Data}\label{sec:data}
\subsection{Residual Level-1B data}
In this study, we use an intermediate GRACE product between Level-1B and Level-2, provided by CSR, denoted here as the residual Level-1B data product. GRACE Level-1B data are derived from the Level-1A data, which are the processed telemetry measurements converted to engineering units obtained from the satellites' instruments. The Level-1A data undergo extensive and irreversible preprocessing, resulting in Level-1B data products, available at 1-5 second sampling rates. 

The residual Level-1B data used in this study consist of CSR's RL06 residual K/Ka-Band Ranging (KBR) Level-1B data for the entire duration of the GRACE mission, which spanned from April 2002 to June 2017. We received both the residual Level-1B data (known as \textit{pre-fit}) and the corresponding Least-Squares fit after the gravity retrieval, limited to degree 180. While these residuals are not publicly available, the original Level-1B datasets, generated by the Jet Propulsion Laboratory (JPL), can be accessed publicly \cite{case2002grace} at \url{https://doi.org/10.5067/GRJPL-L1B03}. Additionally, the processing standards and details developed by CSR \cite{CSR2018grace} to transition from Level-1B to Level-2 are also available to the public.

In this context, we define three essential variables:
\begin{itemize}
    \item \textbf{Pre-fit} ($\delta \dot{\rho}$): Represents the reduced L1B range rate observations, used as input for the unconstrained least-squares gravity inversion to derive Spherical Harmonic (SH) coefficients.

    \item \textbf{Geo-fit} ($\delta \dot{\rho}^{\rm (g)}$): Refers to range rate residuals associated with the fitted pre-fit data, expressible solely in terms of SH coefficients, up to degree and order 180 for this analysis.

    \item \textbf{Post-fit Residuals} ($\Delta_{\delta \dot{\rho}}$): Represents the difference between pre-fit range rates and the updated least-squares fit, known as the post-fit residuals.
\end{itemize}


The following relation is used to summarise what was mentioned above:
\begin{equation}
    \delta \dot{\rho} = \delta \dot{\rho}^{\rm (l)} + \delta \dot{\rho}^{\rm (c)} +\delta \dot{\rho}^{\rm (g)} + \Delta_{\delta \dot{\rho}},
\end{equation}
where $(\rm l)$, $(\rm c)$ and $(\rm g)$ refer to the local, common and global residual range rate components \cite{Pini2012}, and $\Delta_{\delta \dot{\rho}}$ is the post-fit range rate residuals. Global parameters are defined as constants applicable to all arcs and datasets, including the monthly estimated SH coefficients. An arc refers to a continuous orbital trajectory, with each arc associated with three distinct datasets: KBR1B, GRACE-A, and GRACE-B GPS data. Common parameters encompass those shared across all models and arcs, such as the initial state vector, non-gravitational accelerations, and various accelerometer metrics; however, they exclude the monthly SH coefficients. Local parameters, on the other hand, pertain to specific arcs, including aspects like the GPS positions of A and B, KBR phase breaks, and others. For more information, the reader can refer to \citet{Pini2012}. Geo-fits are thus a special category of the global parameters model response, purely written in terms of the monthly SH coefficients. This study primarily focuses on post-fit residuals.

\subsection{GRACE Daily Gravity Solutions}

In this study, we use the ITSG-Grace2018 Kalman-filtered daily solutions \cite{Kvas2019} \footnote{\url{https://ftp.tugraz.at/outgoing/ITSG/GRACE/ITSG-Grace2018/daily_kalman/netcdf/}}, as a crucial component for validation our findings on high-frequency mass change phenomena. These solutions come from the most recent version of ITSG's Kalman-filtered gravity data, which are SH coefficients provided up to degree $l_{\rm max} = 40$ and have a daily temporal resolution. These datasets are also available in gridded netCDF format. Although the spatial resolution is somewhat limited with a (half-wavelength) spatial resolution of 500 km, these gravity field solutions represent the state-of-the-art in terms of temporal resolution. 

\subsection{Level-4 altimetry products}
To enhance our understanding and characterisation of the post-fit residuals related to high-frequency mass change phenomena occurring over the oceans, we analysed gridded Sea-Level Anomalies (SLA) from a Level-4 multi-mission altimetry product provided by the E.U. Copernicus Marine Service Information  \url{https://doi.org/10.48670/moi-00148}. These gridded data are available in netCDF format for the entire period from January 1993 to May 2023, covering the duration of the GRACE mission.

\subsection{Meteorological data}
We also utilise publicly available cyclone tracking and rainfall data from the Bureau of Meteorology (BoM) \url{http://www.bom.gov.au/} for our analysis of meteorological high-frequency mass change phenomena. The cyclone tracking data is in CSV format and includes information on all past cyclones, while the rainfall data is presented in a latitude-longitude gridded format and measured in millimeters.
\section{Methods}\label{sec:methods}
\subsection{Typical acceleration approach}\label{subsec:acc_approach}
As a first step, let us recall the geometry of the GRACE system, which consists of two satellites denoted here as A and B, separated by an approximate distance of $220$ km in the Line-of-Sight (LoS) direction. To derive the Equation of Motion (EoM), an instantaneous relative reference frame (IRRF) is used, with its main directions being LoS, radial, and out-of-plane. 


The range $\rho$, range rate $\dot{\rho}$ and range acceleration $\ddot{\rho}$ can be computed with \cite{Ghobadi_Far2018} 
\begin{equation}
\rho = \mathbf{x}_{AB} \cdot \mathbf{e}_{AB}^{\textrm{LoS}},
\end{equation}
\begin{equation}
\dot{\rho} = \dot{\mathbf{x}}_{AB} \cdot \mathbf{e}_{AB}^{\textrm{LoS}},
\end{equation}
\begin{equation}\label{eq:xAB_dotdot_kinematic}
    \ddot{\mathbf{x}}_{AB} \cdot \mathbf{e}_{AB}^{\textrm{LoS}} = \ddot{\rho} - 
\frac{1}{\rho}(\dot{\mathbf{x}}_{AB} \cdot \dot{\mathbf{x}}_{AB} - \dot{\rho}^2).
\end{equation}
\noindent where $\mathbf{x}_{AB}$, $\dot{\mathbf{x}}_{AB}$ and $\ddot{\mathbf{x}}_{AB}$ are the relative position, velocity and acceleration vectors of the GRACE-A and -B satellite pair (in an Earth-Centered Earth-Fixed frame), and $\mathbf{e}_{AB}^{\rm LoS}$ is the unit vector in LoS direction. The second term on the right-hand side of Equation \ref{eq:xAB_dotdot_kinematic} is typically referred to as the centrifugal acceleration term (which from now is summarised as $g_{\rm c})$.

However, it is evident that the gravity field is still not explicit in the latter equation. This situation can be addressed by using Newton's second law of motion for both A and B, and then taking their difference.
\begin{equation}\label{eq:xAB_dotdot}
    \ddot{\mathbf{x}}_{AB} =   \mathbf{g}^{\rm bkg}_{AB} +
    \delta \mathbf{g}_{AB}+  \mathbf{a}^{\rm ng}_{AB},
\end{equation}
where $\mathbf{a}^{\rm ng}_{AB}$ represents all non-gravitational accelerations, such as air drag and pressure radiation. Meanwhile, $\mathbf{g}^{\rm bkg}_{AB}$ and $\delta \mathbf{g}_{AB}$ denote the gravity differences between satellites GRACE-A and GRACE-B. Specifically, $\mathbf{g}^{\rm bkg}_{AB}$ is associated with all known background models, while $\delta \mathbf{g}_{AB}$ reflects the time-variable gravity signal of interest. 

To derive the residual range rate variables that we use in this study, CSR applied background models related to the static gravity field (GGM05C), the de-aliasing AOD1B product, and other (non-)tidal and non-gravitational accelerations. For more information on the background models and processing strategy implemented by CSR, the reader is referred to \cite{CSR2018grace}. Residual quantities resulting from removing these orbital and background models will be denoted with $\delta$. Thus, by combining both Equations \ref{eq:xAB_dotdot_kinematic} and \ref{eq:xAB_dotdot} (projected onto the LoS direction), the resulting residual relation is
\begin{equation}
    \delta g^{\rm LoS}_{AB} = \delta \ddot{\rho} + \delta g_{\rm c},
\end{equation}
where $\delta g^{\rm LoS}_{AB}$ and $\delta g_{\rm c}$ are the residual Line-of-Sight Gravity Difference (LGD) and centrifugal term, respectively.  

\subsection{From range rate to Line-of-Sight Gravity Difference residuals}
It is important to note that we are provided with residual range rates, denoted as \(\delta \dot{\rho}\). There is no straightforward method to compute \(\delta g_{\rm c}\) without performing the extensive orbital modeling done by CSR. Additionally, one of the goals of this study is to demonstrate the effectiveness of this intermediate GRACE product, which is always produced when generating unconstrained Level-2 gravity solutions. Therefore, we follow three steps to obtain the residual LGDs from the residual range rates.

First, we start by applying a Butterworth low-pass (zero-phase) filter to the residual range rates to eliminate the high-frequency noise introduced by the KBR instrument noise, using a cutoff frequency of 11 mHz. This is necessary because noise begins to dominate the signal at this frequency \cite{Goswami2018, Ghobadi_Far2018, Peidou2022}. Additionally, we need to remove long wavelengths that can obscure geophysical signals at frequencies below 0.9 mHz \cite{Ghobadi_Far2018}. As a result, we also apply a high-pass filter. This combined filtering approach effectively creates a band-pass filter with a characteristic frequency range of [0.9, 11] mHz, roughly corresponding to SH degrees 5 to 60, where geophysical signals are present.

Secondly, we apply a fourth-order accurate central numerical differentiation scheme (with $n=2$), in the form of a convolution of $w_{i,j}$ parameters which can be obtained by an iterative process presented below \cite{Khan1999}:
\begin{equation}
    w_{1, 2n} = \frac{n}{n+1},
\end{equation}
\begin{equation}
    w_{k, 2n} = -w_{k-1, 2n} \frac{(k-1)(n-k+1)}{k(n+k)},\ \ \ \ k = 2, 3, 4, ..., n,  
\end{equation}
\begin{equation}
    w_{-l, 2n} = -w_{l, 2n}, \ \ \ \ l = 1, 2, 3, ..., n.  
\end{equation}
Note the central coefficient is set to $w_{0, 2n} = 0$. At the edges of each data section, a forward and backward scheme is applied to ensure that we do not have any edge-related numerical errors caused by the central scheme. 

We now have at our disposal a band-pass filtered residual range acceleration time-series, $\delta \ddot{\rho}(t)$, which has been widely used as an approximation of LGD (e.g. by \citet{Killett2011, weigelt2017acceleration, Spero2021, Allgeyer2022}), however by removing the centrifugal residual, one can better localise mass anomalies on the Earth's surface \cite{Ghobadi_Far2018}. Thus, as a third and final step, an analytical transfer function, $Z(f)$, is applied on $\delta \ddot{\rho}$ as follows \cite{Ghobadi_Far2018}
\begin{equation}\label{eq:Ghobardi_Far_transfer_func}
    \delta g^{\rm LoS}_{AB}(t) =  \mathcal{F}^{-1}\{Z(f) \mathcal{F}\{ \delta \ddot{\rho}(t)\} \} + \Delta_{g_{\rm c}},
\end{equation}
where $f$ is the frequency, $\mathcal{F}$ is the Fourier transform, $\Delta_{g_c}$ is the transfer function error related to the centrifugal term, and lastly the transfer function analytical fit is \cite{Ghobadi_Far2018}
\begin{equation}
    Z(f) = 1 + 3.45 \times 10^{-4} f^{-1.04}.
\end{equation}
The use of this empirical function leads to an accuracy improvement from 50-70\% compared to the typical residual range acceleration approximation, where the global RMS of the band-passed filtered $\Delta_{g_c}$ was found to be below 0.2 nm/s$^2$ \cite{Ghobadi_Far2018}.  Applying this transfer function is therefore an essential processing step to improve the accuracy of our geophysical analysis.

\subsection{Spatial-temporal Linear Regression}
One of the main contributions of this paper is the characterization of the sub-monthly content in post-fit residuals. For this purpose, we perform a linear temporal parametrization per grid cell of the sub-monthly gravity signal in post-fit residuals. Thus, in each grid cell described by the longitude-latitude pair \((\lambda_0, \theta_0)\), we assume the following relationship:

\begin{equation}\label{eq:lin_reg}
    \delta g^{\rm LoS}_{AB}(\lambda_0, \theta_0, t) \approx a(\lambda_0, \theta_0) t + b(\lambda_0, \theta_0),
\end{equation}
\noindent where \(a\) and \(b\) represent the trend and linear intercept (at time $t=0$), respectively. These parameters are computed using the known regression equations:

\begin{equation}\label{eq:a}
    a = \frac{n \sum_i^n (t_i (\delta g^{\rm LoS}_{AB})_i) - \sum_i^n t_i \sum_i^n (\delta g^{\rm LoS}_{AB})_i}{n \sum_i^n t_i^2 - (\sum_i^n t_i)^2},
\end{equation}

\begin{equation}\label{eq:b}
     b = \frac{\sum_i^n (\delta g^{\rm LoS}_{AB})_i - a \sum_i^n t_i}{n},
\end{equation}
\noindent where \(t_i\) and \((\delta g^{\rm LoS}_{AB})_i\) denote the \(i\)th data points in each grid cell, with \(n\) being the total count of data points.

The results from Equations \ref{eq:a} and \ref{eq:b} yield the spatio-temporal monthly distributions of the sub-monthly gravity field and other non-gravitational residuals, as discussed in previous sections.

\section{Analysis of KBR residual Level-1B data in the context of high-frequency mass change variations}\label{sec:results}
High-frequency mass change phenomena occur across various spatial and temporal scales, driven by processes within the atmosphere, oceans, land hydrology, and the solid Earth. These mass change signals can be observed and characterised using GRACE gravimetry data. In this study, we focus on analysing post-fit residuals in the form of Line-of-Sight Gravity Differences (LGDs). 
\subsection{Global overview of high-frequency mass change phenomena}
In this section, we present an overview of the major categories of high-frequency change signals and characterise their spatial distribution.
\subsubsection{Statistics of post-fit residuals}
To visualise the signal content in post-fit residuals, a spatial analysis is performed. By "spatial," we mean that the data is binned into latitude-longitude grids, and several statistics are computed based on the data points falling within these bins.

\begin{figure}[htpb]
    \centering
    \includegraphics[width=1.05\textwidth]{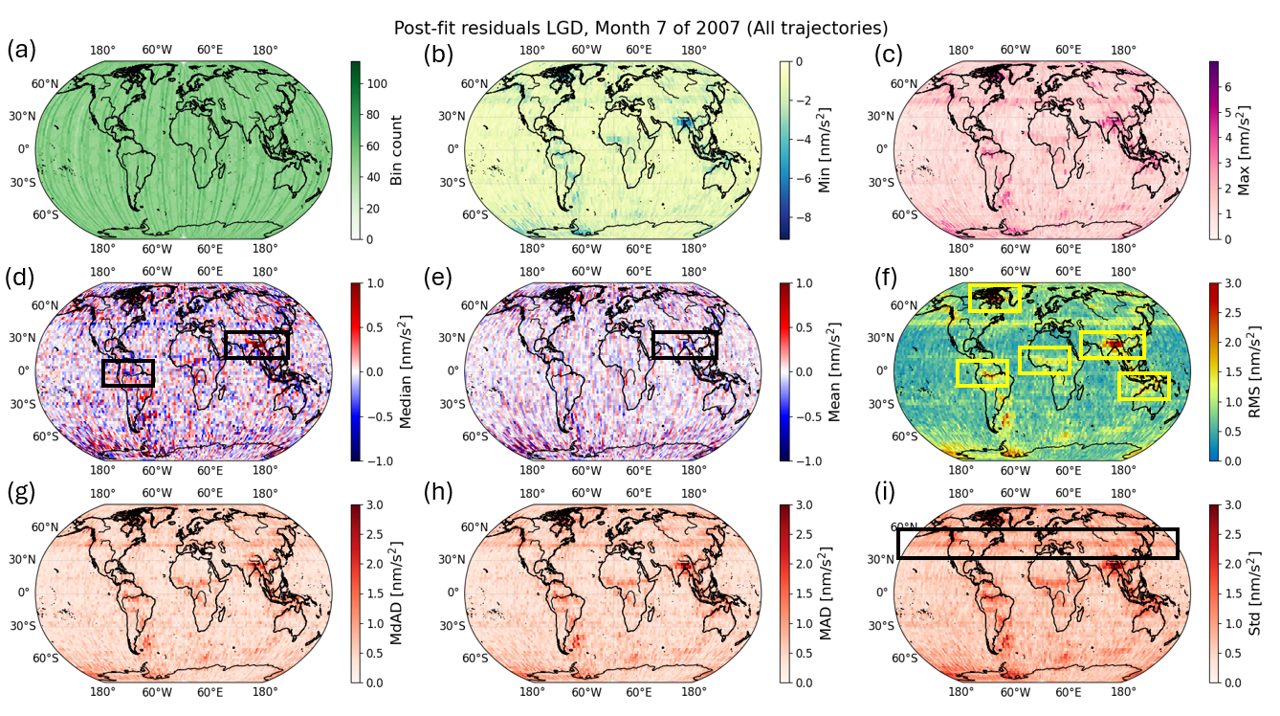}
    \caption{Spatially binned (2.5°x2.5°) statistical layout of Line-of-Sight Gravity Difference post-fit residuals for the month of July 2007. Top row: (a) bin count, (b) minimum, and (c) maximum. Middle row: (d) median, (e) mean, and (f) RMS. Bottom row: (g) Median Absolute Deviation (MdAD), (h) Mean Absolute Deviation (MAD), and (i) standard deviation (std).}
    \label{fig:stats_2007_07}
\end{figure}

\begin{figure}[htpb]
    \centering
    \includegraphics[width=1.05\textwidth]{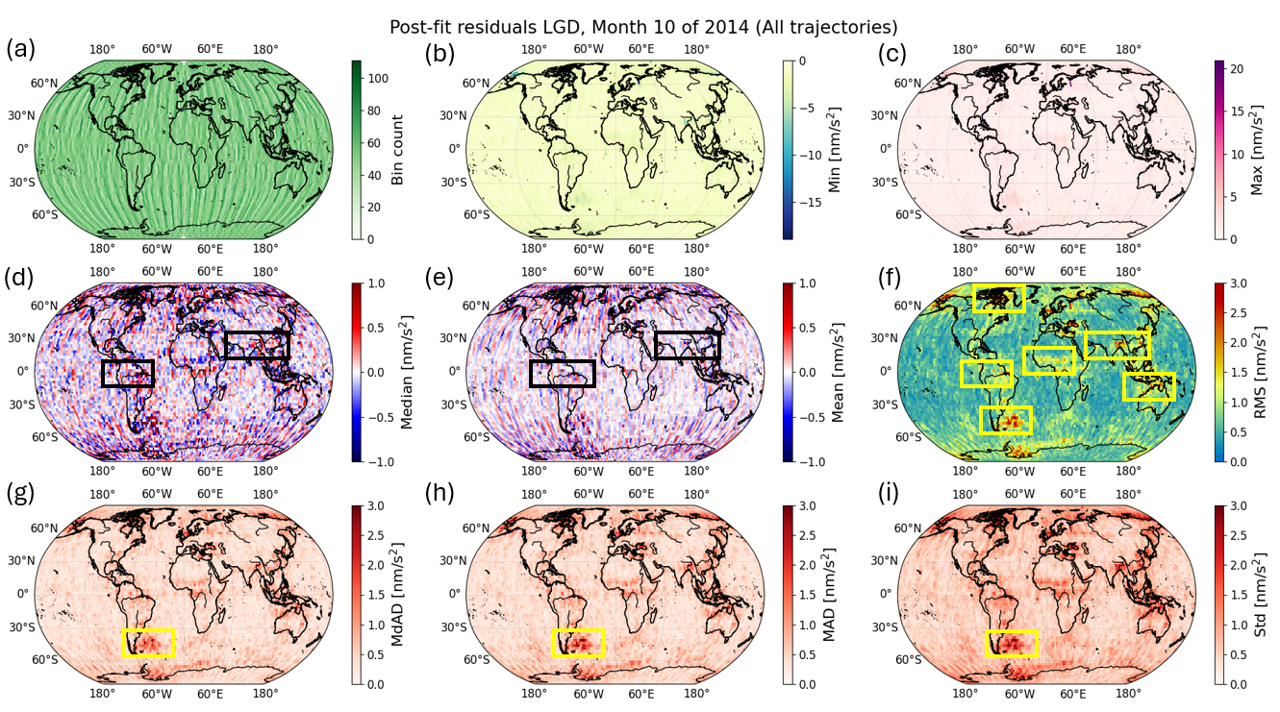}
    \caption{Spatially binned (2.5°x2.5°) statistical layout of Line-of-Sight Gravity Difference post-fit residuals for the month of October 2014. Top row: (a) bin count, (b) minimum, and (c) maximum. Middle row: (d) median, (e) mean, and (f) RMS. Bottom row: (g) Median Absolute Deviation (MdAD), (h) Mean Absolute Deviation (MAD), and (i) standard deviation (std).}
    \label{fig:stats_2014_10}
\end{figure}

We observed that the spatial root mean square (RMS), as well as other statistical variance estimators, such as standard deviation and median/mean absolute deviations (MdAD and MAD respectively), of the post-fit residuals offer an optimal way to visualize the different types of high-frequency signals, as opposed to other statistical measures such as the spatial mean or median. To explain this, two randomly selected monthly statistical layouts of binned GRACE post-fit residuals are presented in Figures \ref{fig:stats_2007_07} (07/2007) and \ref{fig:stats_2014_10} (10/2014).

Two general observations emerge from the plots. First, while the binned mean (Figures \ref{fig:stats_2007_07}e and \ref{fig:stats_2014_10}e) post-fit LGDs show no particular pattern, the median values (Figures \ref{fig:stats_2007_07}d and \ref{fig:stats_2014_10}d) exhibit more structure, especially noticeable in regions such as South Asia, Central Africa, and the Amazon basin. Secondly, the variance plots—including standard deviation, mean absolute deviation, and median absolute deviation—along with the root mean square (RMS) and maximum values (Figures \ref{fig:stats_2007_07}\&\ref{fig:stats_2014_10}b, c, f, and g to i), display structures that correlate with sub-monthly geophysical phenomena and other mis-modelled artefacts. We hypothesise that an example of accelerometer artefact is particularly evident in the post-fit residuals RMS and the variances of 1 nm/s$^2$ observed as a continuous 45°N latitudinal band, as shown in Figure \ref{fig:stats_2007_07}(i). This band cannot be explained by any geophysical phenomenon and as it correlates with the spatial patterns of monthly inter-satellite pointing pitch variations presented by \citet{Bandikova2012}, we hypothesise that these are residual thruster accelerometer signals \cite{Bandikova2019}. These and other accelerometer remnants (i.e. roll, yaw and combined variations) seem to be present throughout the majority of post-fit residuals' plots during the GRACE mission. 

The difference between the median/mean statistics (e.g. Figures \ref{fig:stats_2007_07}d and \ref{fig:stats_2007_07}e), which show little to no geophysical correlation, and the RMS and variances, which do exhibit such correlations, can be explained by the nature of the post-fit residuals. These residuals indicate mass anomalies relative to the monthly solutions (geo-fit), leading them to oscillate around the monthly average, causing them to superimpose and cancel each other out, resulting in a lack of any discernible pattern. In contrast, the amplitude of these oscillations is captured by analysing the power of the signal (RMS) or by examining their absolute or squared deviations (MAD or MdAD). Due to this, the RMS will be used throughout this paper to visualise the power of GRACE's unmodelled reduced observations of sub-monthly signals.

\subsubsection{Temporal characteristics of post-fit residuals}\label{subsec:temp}
\begin{figure}[ht]
    \centering
    \includegraphics[width=0.975\textwidth]{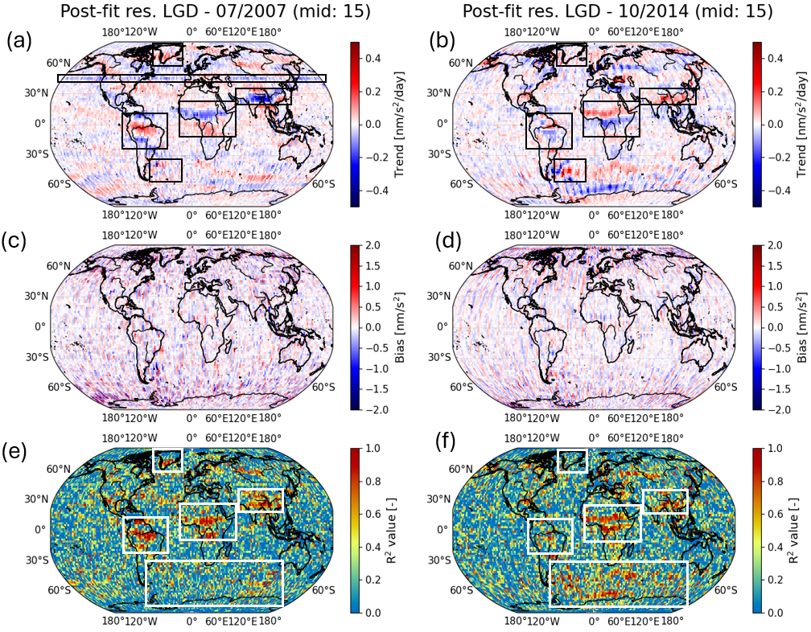}
    \caption{Spatio-temporal linear regression grid (2°x2°) showing Line-of-Sight Gravity Difference post-fit residuals for the month of July 2007 (left) and October 2014 (right). (a) and (b): trend in (nm/s$^2$)/day, (c) and (d): bias at 15 days in nm/s$^2$, and (e) and (f) adjusted R$^2$-value. }
    \label{fig:lin_reg_2007_2014}
\end{figure}

Following the description of the simple binning statistics of the GRACE post-fit residuals, we now present the first temporal characterization for the same two months mentioned earlier. This is illustrated in Figure \ref{fig:lin_reg_2007_2014}, using a temporal linear regression parameterization. It is important to note that we observed a strong inverse correlation, ranging from -0.95 to -0.99, between the slope and intercept. Therefore, instead of presenting the intercept, we provide the "bias," which refers to the interpolated value of LGD at the midpoint of the month, as defined by Equation \ref{eq:lin_reg}.

Several observations can be made from Figure \ref{fig:lin_reg_2007_2014}. First, regions with high spatial RMS, as shown in Figures \ref{fig:stats_2007_07}f and \ref{fig:stats_2014_10}f, correspond directly to areas that exhibit distinct trend signals in Figures \ref{fig:lin_reg_2007_2014}a and \ref{fig:lin_reg_2007_2014}b. Notably active river basins, such as the Amazon and Ganges-Brahmaputra, show trend values of up to ±0.5 nm/s²/day. Additionally, African basins like the Niger, Chad, (South) Nile, and Congo exhibit trend values of ±0.25 nm/s²/day. Furthermore, the mid-month LGD bias presented in Figures \ref{fig:lin_reg_2007_2014}c and \ref{fig:lin_reg_2007_2014}d does not show any discernible geophysical patterns, as the values remain below ±0.5 nm/s². This indicates that the sub-monthly characteristics of post-fit residuals can be sufficiently described by a linear trend alone. Finally, the large trend values result in LGD sub-monthly variations of up to 15-20 nm/s² throughout the month, corresponding to more than half the magnitude of the monthly signals.

Second, these river basins show the highest $R^2$ values (from 0.7 to 0.98), while the northern oceanic regions display low $R^2$ values (below 0.5). Furthermore, the southern oceanic regions demonstrate higher sub-monthly activity than expected \cite{Bonin2020, Schindelegger2021}. This indicates that sub-monthly land hydrology, caused by sudden events such as floods and heavy precipitation, can be identified and characterized using spatially distributed linear regression. In contrast, regions of the ocean with low $R^2$ may be affected by residual KBR noise \cite{Peidou2022}.

Thirdly, this methodology also allows us to identify unmodeled AOD1B and ocean tidal signals related to ocean currents in Figure \ref{fig:lin_reg_2007_2014}b and f, particularly related to the sub-tropical gyre in the Argentine basin. Finally, we can distinguish the positive sub-monthly trend of 0.1 nm/s$^2$/day in southern Greenland during the summer month of July 2007 (which is not observed during the autumn month of October 2014 in Figure \ref{fig:lin_reg_2007_2014}b), as well as the hypothesized accelerometer remnant with a trend of -0.08 nm/s$^2$/day along the 45°S latitudinal band in Figure \ref{fig:lin_reg_2007_2014}a. These results related to the linear regression of post-fit residuals suggest that it is possible to immediately improve the high-frequency resolution of monthly models.

\begin{figure}[ht]
    \centering
    \includegraphics[width=0.95\textwidth]{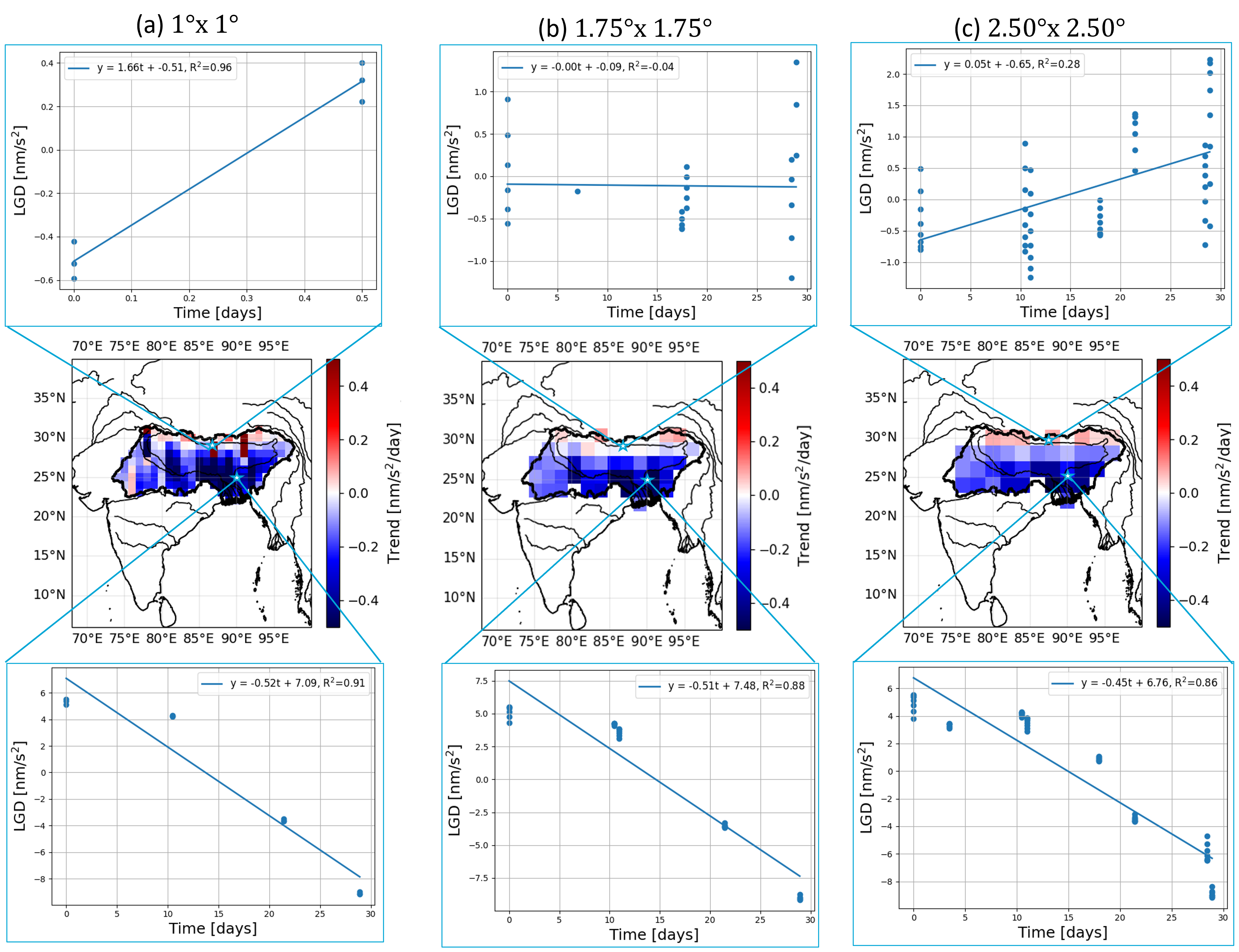}
    \caption{Linear regression trends depicted in a grid format for three distinct grid sizes, analyzing time series data for two specific grid cells during July 2007. The grid sizes include: (a) 1° x 1°, (b) 1.75° x 1.75°, and (c) 2.5° x 2.5°. This analysis illustrates variations in trends across different spatial resolutions.}
    \label{fig:grid_time_series}
\end{figure}

Finally, we verify our spatiotemporal linear regression strategy by focusing on the time series of two grid cells for July 2007 within the Ganges-Brahmaputra river basin in Figure \ref{fig:grid_time_series}.

In the top and last row of Figure \ref{fig:grid_time_series}, it is evident that increasing the grid size results in a greater number of data points and their corresponding arc clusters. In areas with a clear trend, enlarging the grid cell size (as shown in Figures \ref{fig:grid_time_series}a to c) does not significantly impact the final trend estimate. However, several large artefacts are noticeable, particularly in the top row. These artefacts arise from the short timeframe over which the data is collected within the grid cell, leading to a high trend value of 1.66 nm/s²/day (see the top row of Figure \ref{fig:grid_time_series}a). As the grid cell size increases, this outlier diminishes, providing a more accurate representation of the signal trend. For this study, we have chosen a grid cell size of 1.75° for both latitude and longitude when discussing trend values. The choice of grid cell size and mesh optimization is considered beyond the scope of this work.

\subsubsection{Sub-monthly geophysical signals}
Our general analysis of all monthly GRACE post-fit residuals solutions (such as Figures \ref{fig:stats_2007_07} to \ref{fig:lin_reg_2007_2014}) reveals two primary categories of high-frequency signals, which is in strong agreement with previous studies \cite{Ghobadi_Far2020, Peidou2022, Ghobardi_Far2022}:
\begin{enumerate}
    \item \textbf{Hydro-meteorological and Oceanic processes:} Non-modelled phenomena in background AOD1B and ocean tidal products, particularly in terms of atmospheric and oceanic mass change, is a significant source of high-frequency signals \cite{Bonin2020, Schindelegger2021, Ghobardi_Far2022}. This is especially evident in the global oceans, where residuals indicate unmodeled effects from short-term variability in oceanic currents and pressure systems, such as the mass anomalies related to the Zapiola region in the Argentine basin as can be seen highlighted from the RMS (and other variance plots) post-fit residuals in Figures \ref{fig:stats_2014_10}f to \ref{fig:stats_2014_10}i.
    \item \textbf{Land Hydrology:} Sub-monthly land hydrology signals are particularly prominent in regions with large river basins, such as the Amazon, central Africa and South Asia (see in particular the large variabilities in Bangladesh and India highlighted in Figure \ref{fig:stats_2007_07}f). Here, rapid changes in water storage due to precipitation and river discharge (e.g. during monsoon seasons) contribute to the high-frequency mass change.
\end{enumerate}
In this study, we only focus our analysis on these two general categories. Solid Earth processes, specifically the resulting tsunami loadings after the December 2004 Sumatran and March 2011 Tohoku earthquakes were observed in LGD residuals \cite{Ghobadi_Far2020}. Lastly, glaciology is also regarded as outside the scope of this study. To date, we have not found any research on the sub-monthly characteristics of LGD or range-rate residuals in polar regions. This absence of studies may be due to the fact that glaciology involves relatively slower geophysical processes, mostly characterised by seasonal variations. Both categories will be considered for future research.


Our findings support the results of the previously mentioned studies. However, we significantly expand upon these results in the following sections by providing a detailed spatial and temporal characterisation of specific examples of rapid mass change phenomena throughout the entire duration of the GRACE mission.


\subsection{Oceanic high-frequency mass change signals}\label{subsec:oceanic}
The Zapiola region, situated within the Argentine Basin in the South Atlantic Ocean off the coast of Argentina (centered at 44°S 45°W), is a unique and dynamically active area characterised by intense oceanic currents and mass change. This region hosts the Zapiola Anticyclone, a large, clockwise-rotating, deep-reaching current system that forms over the Zapiola Rise—an elevated underwater plateau. The anticyclonic circulation around the Zapiola Rise traps water masses and sediment, leading to highly concentrated mass anomalies \cite{deMiranda1999, Volkov2008_zap}.

In terms of high-frequency mass change, the Zapiola region is significant due to the strong currents and complex interactions between the water column and seabed, which create rapid and localised mass variations. These fluctuations produce measurable, sub-monthly signals in gravimetric data, such as a dipole-like barotropic mode which rotates with a 20-25 days period already detected by \citet{Han2014_zap, Yu2018_zap} using 10-day GRACE solutions \cite{Lemoine2007}. Furthermore, \citet{Ghobardi_Far2022} showed that localised anomalies detected by altimetry, in the form of Sea Level Anomalies (SLA), could explain along-orbit LGD residual variations in Laser Ranging Interferometry (LRI) L1B observations. 

\begin{figure}[htpb]
    \centering    \includegraphics[width=1\textwidth]{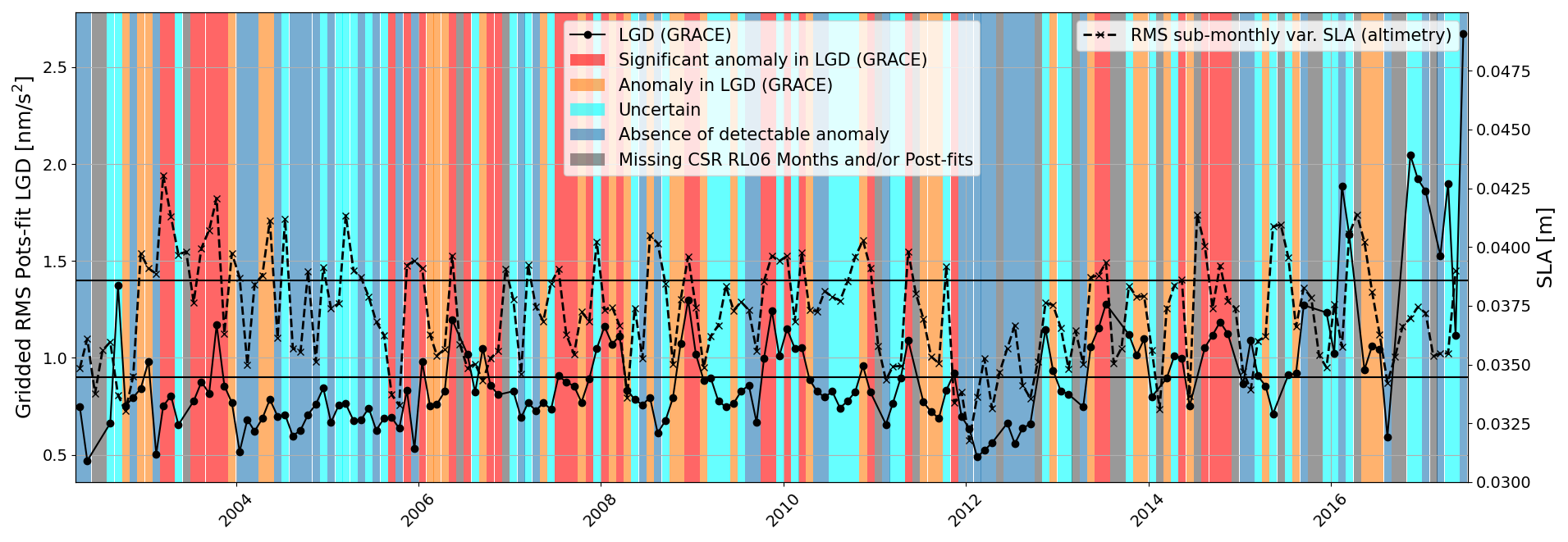}
    \caption{Time series of gridded RMS LGD post-fit residuals, and RMS sub-monthly variability of daily gridded Sea Level Anomalies (SLA) in the Zapiola region for the entire duration of the GRACE mission. A qualitative assessment, indicated by colour bands into four categories: absence (blue), uncertain (cyan), anomaly (orange), and significant anomaly (red), is shown to determine whether the observed signals are due to high-frequency mass variations. Finally, the threshold range of [0.92, 1.42] nm/s$^2$ is illustrated as solid horizontal lines.}
    \label{fig:rms_time_func}
\end{figure}

We conduct the first high-level characterisation of the oceanic gyre high-frequency signals by calculating the monthly mean of the spatially gridded RMS post-fit residuals over the region defined by [35°S-55°S; 27.5°W-55°W] for the entire duration of the GRACE mission, as shown in Figure \ref{fig:rms_time_func}. Each point in Figure \ref{fig:rms_time_func} represents the magnitude of the unmodelled geophysical high-frequency signals, which has its most likely source in errors of the AOD1B and other background models, with the addition of other artefacts such as initial state vector, and accelerometer errors. Furthermore, we performed a qualitative assessment, through a heuristic visual inspection approach of all monthly post-fit residual plots (which are provided in the SI), which are represented by arbitrarily chosen colour bands that range from significant anomalies to uncertain variations, and even the absence of anomalies. This approach allows us to consider that a post-fit residual signal RMS may also represent other sources other than unmodelled AOD1B, such as various accelerometer artefacts (see previous section). The observed increase in RMS since mid-2016 corresponds with the final GRACE era, during which the satellites relied on only one accelerometer.

\begin{figure}[htb]
    \centering
    \includegraphics[width=1\textwidth]{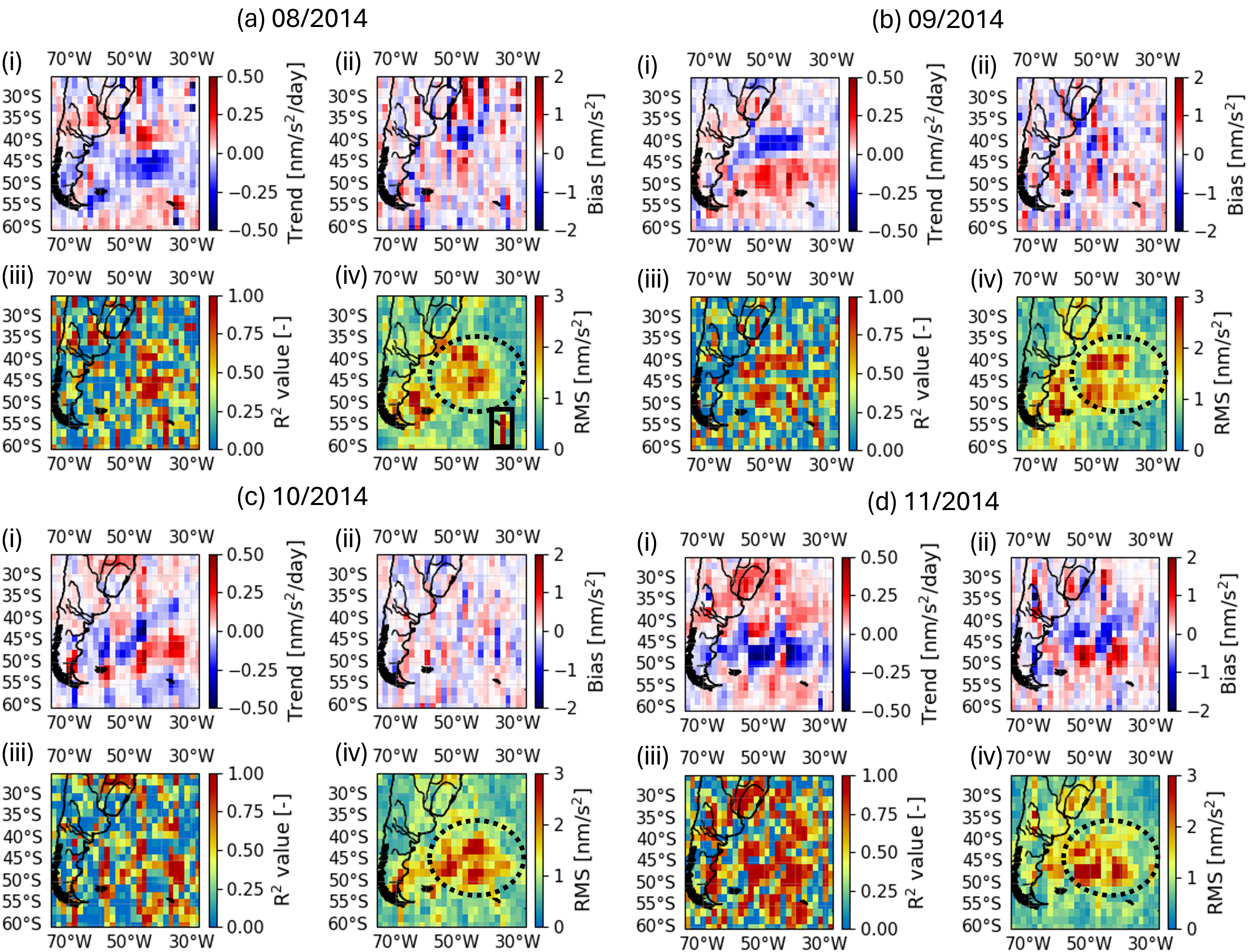}
    \caption{Four consecutive monthly grid groupings displaying spatially binned (1.75° x 1.75°) Line-of-Sight Gravity Differences post-fit residuals and their linear regression results for the months of August (\textbf{a}) to November 2014 (\textbf{d}) in the Zapiola region with sub-plots (i): trend, (ii): mid-month bias, (iii): adjusted $R^2$-value, and (iv): spatially binned RMS.}
    \label{fig:2014_sols}
\end{figure}

We found that within a mean gridded RMS interval of [0.92, 1.42] nm/s², the detection rate is 74\%. This indicates the presence of a consistent geophysical signal during that month. The detection rate is calculated by counting the number of months classified as a "(significant) anomaly" for which the RMS falls within the specified interval. This interval was established by maximizing the detection rate while minimizing the occurrences of no anomalies detected. The associated probabilities for the classification being labelled as "uncertain" and for a lack of detection are 21.7\% and 4.3\%, respectively. The associated statistics of this classification analysis can be found in the Supplementary Information (SI) \footnote{SI can be requested from the corresponding first author.} (Figure S1). It is important to note that, despite the limitations inherent in the visual qualitative assessment, this analysis intends to be a preliminary step towards contextualising the magnitude of mean RMS post-fit residuals. This is the case of the continuous time-series of sub-monthly variabilities detected using this approach of which spatial maps can be found in Figure \ref{fig:2014_sols}, covering the period from August to November 2014 (see the peak categorised as "significant anomaly" in Figure \ref{fig:rms_time_func}).

Figure \ref{fig:2014_sols} shows tidal mis-modellings, which are typically observed in shallow waters, near the coast within the Patagonian shelf (from 45°S to 55°S). Additionally, constant longitudinal artefacts such as the large deviations caused by the mis-modelled initial state vector from Figure \ref{fig:2014_sols}a(i) in August 2014 have been highlighted on the southeastern side. Most significantly, the evolution of sub-monthly mass change is primarily characterised by a rotating pair of negative-positive trend mass anomalies in all (i) sub-plots of Figure \ref{fig:2014_sols}. These reach values similar to land hydrology up to $\pm 0.5$ nm/s$^2$/day. Furthermore, apart from November 2014 (Figure \ref{fig:2014_sols}d(ii)), the mid-month bias does not show any significant geophysical signal. These anomalies appear to rotate around the Zapiola Rise and be constrained by the topography of the Argentine basin. Consequently, we interpret that they are linked to high-frequency components of the known dipole-like rotating mode \cite{deMiranda1999, Yu2018_zap}.

Lastly, we compared this RMS time-series with the sub-monthly variations of Level-4 SLA as provided by the E.U. Copernicus Marine Service Information. The RMS sub-monthly variability is calculated using the following equation:
\begin{equation}\label{eq:RMS_var}
    RMS_{\rm var} = \sqrt{\frac{1}{N}\sum_{i=1}^N(x_i - \bar{x})^2},
\end{equation}
where \( N \) represents the number of gridded daily SLA solutions \( x_i \) within a specific month, and \( \bar{x} \) is the average of all these solutions for that month. The result of Equation \ref{eq:RMS_var} is a grid that depicts the RMS variability of the gridded variable \( x_i \) in relation to its monthly average. The resulting spatial maps from August to November 2014 can be found in the SI (Figures S2) and the respective time-series is plotted with the post-fit residuals in Figure \ref{fig:rms_time_func}.

Our comparison revealed no significant temporal correlation coefficients (below 0.25) between the post-fit residuals LGD RMS and the sub-monthly RMS SLA variability time-series (as can be visually observed in Figure \ref{fig:rms_time_func}). To assess the robustness of these results, we recalculated the correlations after excluding all data prior to June 2003 and after August 2016, in order to eliminate potential issues stemming from the initial problematic periods when one SCA (Star Camera Assembly) was producing inaccurate results and when only one accelerometer was collecting measurements, respectively. Additionally, we segregated the analysis by orbital track type, examining ascending and descending tracks separately. The exclusion of data provided an overall increase of 0.1 (or 50\%). In contrast, the adjustments related to orbital track type led to significant variations from only 0.05 (20\%) to even 0.15 (75\%) in the correlation coefficients, with a maximum value of 0.31 (for ascending tracks). Due to the high sensitivity of low correlation coefficients, spatially averaged SLA variations alone do not adequately explain the high-frequency signals observed by GRACE. This highlights the limitations of such monthly aggregate statistical analyses. As noted by \citet{Ghobardi_Far2022}, for specific ground tracks, the along-orbit LGD residuals exhibited (anti-)correlations with daily SLA variations. Furthermore, it is known that the barystatic (mass-related) component of SLAs dominates the region, while the steric component—driven by variations in temperature and salinity—contributes less than 10\% to the overall regional signal \cite{Hughes2007}.  Consequently, this suggests that other underlying phenomena, such as rapid pressure fluctuations and circulation variations \cite{Ghobardi_Far2022}, may be occurring, which appear to be diminished by the spatiotemporal averaging process.


\subsection{Hydro-meteorological high-frequency mass change signals}\label{subsec:cyclones}
We focus now on the Gulf of Carpentaria (GoC), a shallow sea situated in northern Australia, as our primary case study to examine abrupt hydro-meteorological changes that can be measured using GRACE residual L1B data. This sea is characterised by significant annual variations, with a measurable change in sea surface height of up to 0.4 metres, as detected by GRACE Level-2 data \cite{Tregoning2008}. In addition to these annual and seasonal fluctuations, the region exhibits considerable levels of sub-monthly variability \cite{Schindelegger2021, Ghobardi_Far2022}, which appear to correlate with pressure variations resulting from cyclonic activity \cite{Ghobardi_Far2022}.

During the GRACE period, Cyclone Oswald emerged in the Gulf of Carpentaria on 17 January 2013, leading to considerable flooding in the Northern Territory and Queensland along its western coast by 29 January \cite{bom_oswald}. In this study, we investigate this event by analysing post-fit residuals to offer a gravimetric characterisation of the sub-monthly variations induced by the cyclone's occurrence. 

\subsubsection{Along-orbit analysis}
\begin{figure}[htpb]
    \centering
    \includegraphics[width=1\textwidth]{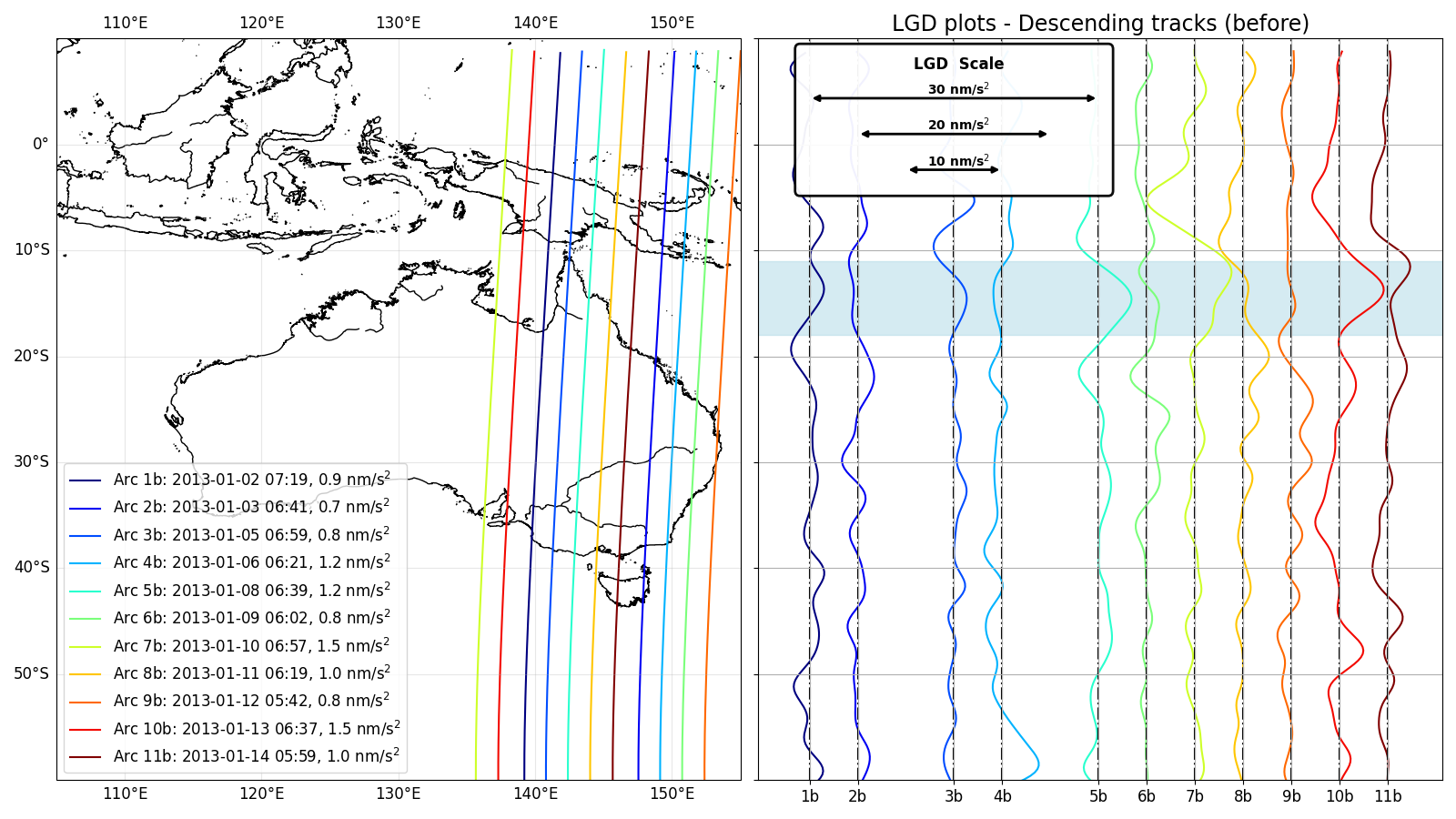}
    \caption{\textbf{Left}: Ground tracks from the GRACE mission, descending arcs (1b to 11b), displayed alongside their corresponding initial dates (prior to Cyclone Oswald) and the associated Line-of-Sight Gravity Difference (LGD) RMS values.  
    \textbf{Right}: Along-orbit LGD post-fit residuals plotted against latitude for the indicated arcs (1b to 11b), with a LGD scale included. The spacing along the x-axis is proportional to the time interval between each consecutive track.}
    \label{fig:lgd_desc_oswald_bef}
\end{figure}

\begin{figure}[htpb]
    \centering
    \includegraphics[width=1\textwidth]{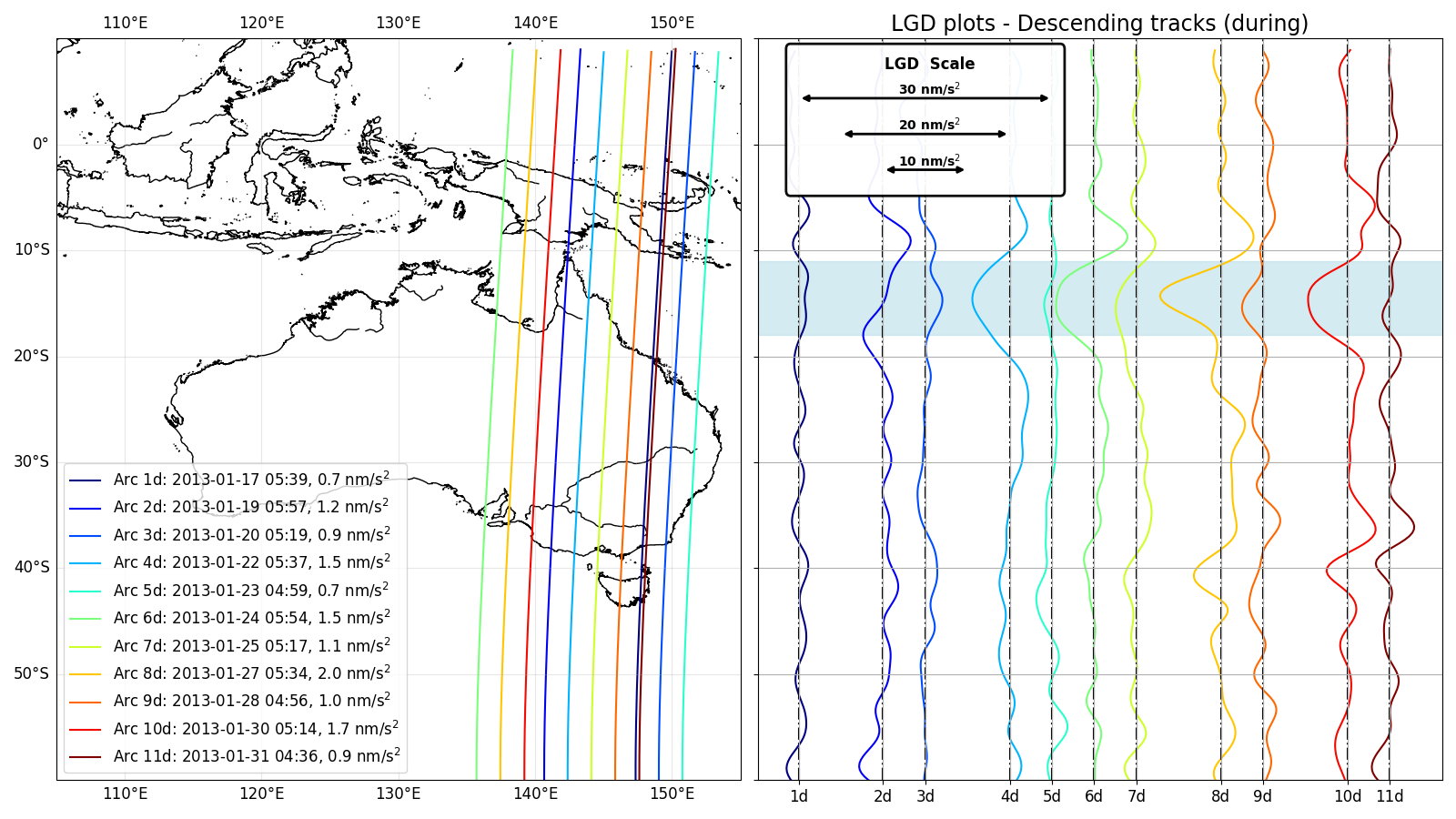}
    \caption{\textbf{Left}: Ground tracks from the GRACE mission, descending arcs (1d to 11d), displayed alongside their initial corresponding dates (during Cyclone Oswald) and the associated Line-of-Sight Gravity Difference (LGD) RMS values.  
    \textbf{Right}: Along-orbit LGD post-fit residuals plotted against latitude for the indicated arcs (1d to 11d), with a LGD scale included. The spacing along the x-axis is proportional to the time interval between each consecutive track.}
    \label{fig:lgd_desc_oswald_aft}
\end{figure}

We start by presenting the (arbitrarily chosen) descending along-orbit variations of the post-fit residuals as a function of latitude next to their ground-tracks, which is a visualisation similar to previous research \cite{Han2021_PNAS, Ghobardi_Far2022}, and is illustrated in Figures \ref{fig:lgd_desc_oswald_bef} and \ref{fig:lgd_desc_oswald_aft}, approx. two weeks before and during Cyclone Oswald respectively. The letters ‘b’ and ‘d’ are appended to the arc indices to differentiate between the "before" and "during" tracks, respectively.


We first focus on the GoC sea near the Northern Territory (where Oswald originated from), where GRACE passed over during arcs 1b, 7b, and 10b before Cyclone Oswald, as illustrated in Figure \ref{fig:lgd_desc_oswald_bef}. In contrast, during the cyclone's occurrence, GRACE directly traversed above it in arcs 6d, 8d, and 10d, shown in Figure \ref{fig:lgd_desc_oswald_aft}. Notably, the arc RMS values observed during Cyclone Oswald, at 1.5, 1.7, and 2 nm/s$^2$, are higher than those recorded before (ranging from 0.9 to 1.5 nm/s$^2$), with arcs 8d and 10d exhibiting the highest RMS values, reaching up to 2 nm/s$^2$. These two arcs exhibit negative LGD variations which peak at 15°S latitude of -7.5 nm/s$^2$ and -5 nm/s$^2$, respectively. The negative values suggest the presence of sub-monthly positive mass anomalies resulting from pressure differences during Oswald, which were not accounted for in the AOD1B RL06 background model \cite{Ghobardi_Far2022}. At similar latitudes (12°S-17.5°S), positive LGD deviations are noted for arcs 7b and 10b, indicating a mass deficit relative to the monthly average. This demonstrates that the monthly solutions have captured a portion of the mass (pressure) variation signals induced by Oswald, revealing an overestimation in mass prior to the cyclone's impact as indicated by the post-fit residuals. 

We conducted a parallel analysis for the ascending tracks and observed comparable results; specifically, the transition from slightly positive to substantially negative LGD variations, reaching up to -5 nm/s$^2$, occurred from the period preceding Oswald to during its presence. 
These observations underscore GRACE's capability to detect instantaneous mass deficits/excess w.r.t. the monthly average associated with extreme weather events, consistent with the prior study by \citet{Ghobardi_Far2022} that highlights its sensitivity to transient mass redistribution caused by atmospheric pressure changes.

\subsubsection{Spatial analysis}
\begin{figure}[htpb]
    \centering
    \includegraphics[width=1\textwidth]{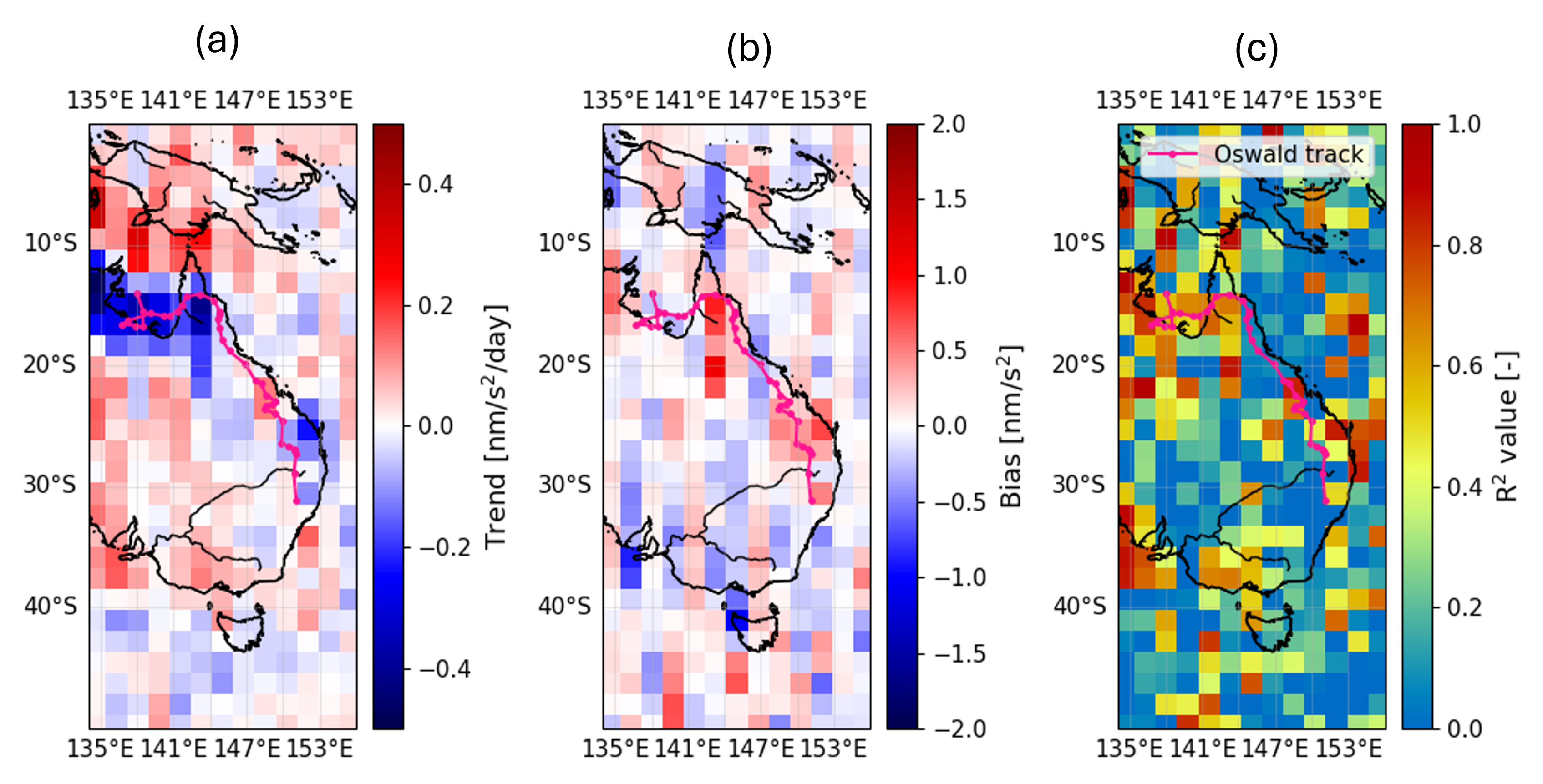}
    \caption{LGD post-fit residuals gridded linear regression (1.75° x 1.75°), with \textbf{a}: Trend in (nm/s$^2$)/day,  \textbf{b}: Mid-month bias (nm/s$^2$), and \textbf{c}: Adjusted $R^2$-value.}
    \label{fig:australia_lin_reg}
\end{figure}

\begin{figure}[htpb]
    \centering
    \includegraphics[width=1\textwidth]{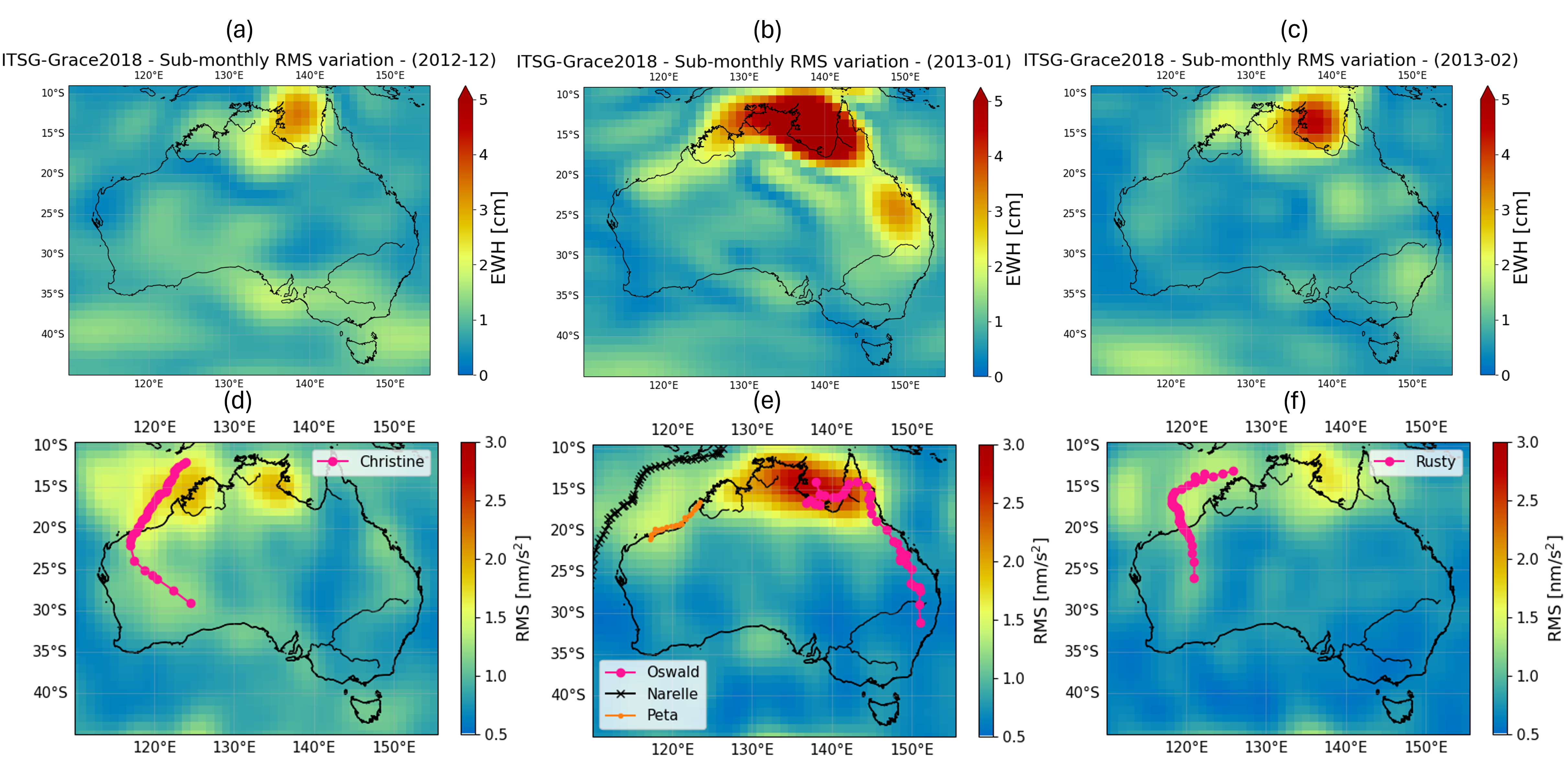}
    \caption{Consecutive monthly grids for December 2012 to February 2013 (left to right) in Australia. The top row (\textbf{a} to \textbf{c}) presents the sub-monthly variations in Equivalent Water Heights (EWH) in cm of daily Kalman filtered solutions from ITSG-Grace2018, and the bottom row (\textbf{d} to \textbf{f}) shows the gridded and spatially smoothed (Gaussian filter with radius 6 deg and $\sigma=2$) LGD post-fit residuals RMS (1°x1°).}
    \label{fig:spatial_maps_dec_to_feb}
\end{figure}

After examining individual trajectories over the GoC, we will now analyse the overall spatio-temporal variations using linear regression on the post-fit residuals, as illustrated in Figure \ref{fig:australia_lin_reg}. In Figure \ref{fig:australia_lin_reg}a, we observe a significant LGD trend ranging from -0.1 to -0.3 nm/s²/day in the GoC and Northern Australia, which corresponds with the cyclone's track and the most significant pressure variations. This finding aligns with our along-orbit results, where we noted a shift from positive to negative instantaneous LGD variations. In contrast to land hydrology (see Section \ref{subsec:temp}), this region with a negative trend has lower \(R^2\) values of only 0.6 to 0.75. This indicates that temporal linear regression alone is insufficient to adequately characterize hydro-meteorological phenomena. Furthermore, while the spatial root mean square (RMS) in Figure \ref{fig:spatial_maps_dec_to_feb}e does not reveal specific patterns at the end of the cyclone's tracks on the eastern coast, the trend does show a small area of negative trend centred at (26°S, 152°E), coinciding with flooding events, and has adjusted \(R^2\) values ranging from 0.5 to 0.7.

To verify that these signals are related to sub-monthly geophysical phenomena, we compare spatially gridded and Gaussian filtered (radius of 6 deg and standard deviation $\sigma=2$) post-fit residuals, from December 2012 to February 2013, with sub-monthly RMS variations (using Equation \ref{eq:RMS_var}) in the ITSG-Grace2018 daily solutions \cite{Mayer_Gurr_T2018,Kvas2019} resulting in Figure \ref{fig:spatial_maps_dec_to_feb}. 
First, note the range for post-fit residuals was set to 0.5 nm/s$^2$ to differentiate actual sub-monthly signal from KBR noise \cite{Peidou2022}. We can initially ascertain that January 2013 (Figures \ref{fig:spatial_maps_dec_to_feb}b and \ref{fig:spatial_maps_dec_to_feb}e) is notably distinctive, exhibiting significant variations over the entire Northern Australia reaching above 3 nm/s$^2$ and above 5 cm EWH in post-fit residuals and sub-monthly RMS, respectively. These variations are particularly evident in the GoC, Queensland, and the Northern Territory (or
$[11^\circ \rm S, 19^\circ S, 132^\circ E, 145^\circ E]$) as during January, Australia experienced three major cyclones. In contrast, both daily solutions and post-fit residuals for December 2012 (Figures \ref{fig:spatial_maps_dec_to_feb}a and \ref{fig:spatial_maps_dec_to_feb}d), as well as for February 2013 (Figures \ref{fig:spatial_maps_dec_to_feb}c and \ref{fig:spatial_maps_dec_to_feb}e), exhibit lower levels of variability within the region with only one cyclone (Christine in December and Rusty in February). This confirms the hypothesis that GRACE detected sub-monthly variabilities attributable to cyclone Oswald, as these regional anomalies align with the cyclone's initial trajectory.

Furthermore, we observe stronger signals in both the sub-monthly RMS and post-fit residuals along the western coast. These variations (which are considerably less discernible in the post-fit residuals, reaching values of only 1.5 nm/s$^2$) appear to correlate with the tracks of the severe tropical cyclone Narelle and Peta, which occurred between the 5th and 15th, and 22nd and 23rd of January, respectively \cite{bom_narelle}. However, although no corresponding signals in the post-fit residuals are detected related to the significant rainfall events on the eastern coast (Figure \ref{fig:spatial_maps_dec_to_feb}e), the daily solutions seem to capture them (Figure \ref{fig:spatial_maps_dec_to_feb}b) with an RMS variation of 2.5-3 cm EWH. Interestingly, the effects of cyclones Christine and Rusty are clearly visible in the post-fit residuals (Figure \ref{fig:spatial_maps_dec_to_feb}d and f), but are not discernable in the ITSG daily solutions (Figures \ref{fig:spatial_maps_dec_to_feb}a and c).

By demonstrating spatial and temporal correlations between both methods, particularly in Northern Australia, where Cyclone Oswald originated, we confirm that post-fit residuals encompass high-frequency geophysical signals induced by hydro-meteorological phenomena. Additionally, our analysis extended to the severe cyclones in February to March 2007 (cyclones George, Jacob and Kara) revealing significant LGD variations—most notably for Cyclone George—which correlated with each cyclone's trajectory (see Figures S4 in the SI). 


\subsubsection{Point-mass model analysis}
\begin{figure}[htpb]
    \centering
    \includegraphics[width=1\textwidth]{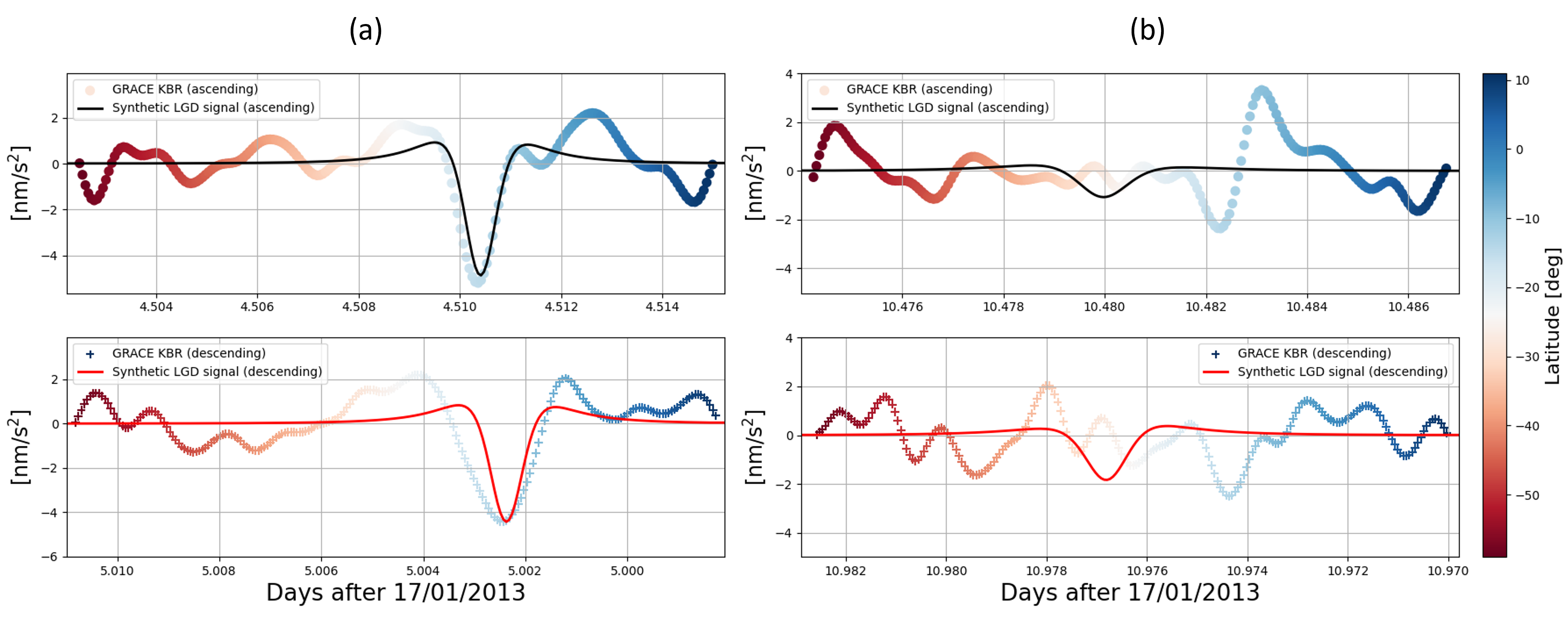}
    \caption{Two examples of along-orbit post-fit LGD residuals (denoted as "GRACE KBR"), compared to synthetic LGDs from a 30 Gt-point mass model of the Oswald cyclone (solid line). The bottom row shows descending tracks (the time-axis is inverted), whereas the top row presents ascending tracks. \textbf{(a)}: Example of alignment between synthetic and post-fit residuals LGDs.  
    \textbf{(b)}: Example of complete misalignment between synthetic and post-fit residuals LGDs.}
    \label{fig:alignment_no_alignment}
\end{figure}

As a final step, we investigate whether these anomalies can directly be attributed to the moving cyclonic system. We assume the cyclone to be a point mass which tries to account for the unmodelled (AOD1B) cyclonic traces in post-fit residuals. We set this mass to be constant and equal to 30 Gt, which is a first-order estimation using spatially integrated weekly precipitation data provided by the Bureau of Meteorology (BoM), for the week of 19-25 of January (for more details see Figure S3 in SI). This mass also coincides with the integrated mass anomaly considering the cyclone's highest pressure difference variations of approx. 10-15 hPa and a radius of 500 km \cite{Deng2020_Oswald}. We limit the analysis to a case in which we see agreement between the model and post-fit residuals in Figure \ref{fig:alignment_no_alignment}a, and one case where there is no alignment in Figure \ref{fig:alignment_no_alignment}b. 

In Figure \ref{fig:alignment_no_alignment}a, one can see that the anomaly is first highly localised as seen from the ascending track with an LGD of -5 nm/s$^2$, and after half a day, the anomaly spreads leading to a wider LGD anomaly with slightly smaller magnitude of -4 nm/s$^2$ in the descending track. This descending track corresponds to arc 4d in Figure \ref{fig:lgd_desc_oswald_aft}, which passes over the northern region of Queensland at the same latitudes as the GoC. In this aligned scenario, the location and timing of both anomalies appear to correlate with the cyclone's path. This suggests that the observed signal is likely tied to the cyclone's highest central pressure drop from 996 to 990-991 hPa, which occured between the 21\textsuperscript{st} and 22\textsuperscript{nd} of January \cite{Deng2020_Oswald}. Furthermore, we can discern information regarding the anomalies' spatial distribution by observing the changes in the width of the LGD peak \cite{Spero2021}. The width of the anomaly matches a point-like mass (below $300$ km) change in the earlier ascending track, whereas, after half a day, the width has increased and no longer matches with the point-mass model suggesting a significant spatial spread (larger than $400$ km as described by \citet{Spero2021}).

Figure \ref{fig:alignment_no_alignment}b demonstrates that there is no alignment observed in both the ascending and descending tracks (with the latter corresponding to arc 8d in Figure \ref{fig:lgd_desc_oswald_aft}). This results in a potential phase delay of 5 minutes, accompanied by post-fit residual anomalies at -15° latitude. The lack of alignment suggests that the observed post-fit residuals cannot be adequately explained by the simple 30 Gt point-mass model, and thus we cannot confirm if the observed post-fit residuals are caused by the moving cyclonic system. This is most likely due to the fact that on the 27th-28th of January, the cyclone started dissipating \cite{bom_oswald}, and thus the pressure anomaly started decreasing slowly again \cite{Leroux2019, Deng2020_Oswald}.

It is important to highlight that the majority of cases we examined displayed only partial or no alignment between the 30 Gt point-mass cyclone model and the post-fit residuals. When we refer to partial alignment, we identify two main issues. 
First, there might be a phase delay in which the model does not match the timing of the post-fit residuals, especially on land. This discrepancy arises potentially because the sub-monthly anomalies in the post-fit residuals represent hydrological phenomena, such as rainfall and runoff, with a smaller influence from the cyclone's pressure variations.
The second issue involves the magnitude of the LGD not aligning with the point mass model. This misalignment suggests problems with the assumption of constant (point) mass, stemming from variations in precipitation and pressure along the cyclone's landfall. This is consistent with the fact that as the cyclone moved inland, its pressure difference decreased from its maximum of 15 hPa (January 22\textsuperscript{nd}) to approx. 2 hPa  (from January 28\textsuperscript{th}) \cite{Leroux2019}. The latter issue could also be attributed to the absence of an observed intersection between the GRACE and cyclone ground tracks. As GRACE is not in proximity to the cyclone's path, the presence of unrelated mass anomalies could superimpose with the cyclone's (un)modelled signal leading to the observed lack of alignment.

The findings presented highlight that there is strong evidence supporting a connection between the cyclone’s path and the observed anomaly in the aligned scenario. This is particularly evident in the tracks from January 21\textsuperscript{st} and 22\textsuperscript{nd}, where the observed anomalies strongly correlate with the cyclone’s movement, timing, and highest pressure variations (Figure \ref{fig:alignment_no_alignment}a). The significant magnitude of these anomalies, which are absent in monthly gravity field models, suggests that the cyclone's intense central pressure drop and rainfall mass redistribution are likely the primary drivers. However, the majority of cases display partial or no alignment, highlighting the limitations of the simple 30 Gt point-mass model in capturing the complexities of the observed signal.



\subsection{Land hydrology high-frequency mass change signals}
The Ganges-Brahmaputra river basin is renowned for its significant hydrological variations, primarily due to the rainy season coinciding with the summer monsoon, which typically commences in June and concludes in mid-October. From June to September 2007, Bangladesh experienced severe flooding, impacting 42\% of the country \cite{Islam2010}, due to the increased melting of the Himalayan glaciers. 

\begin{figure}[htbp]
    \centering
    \includegraphics[width=0.9\textwidth]{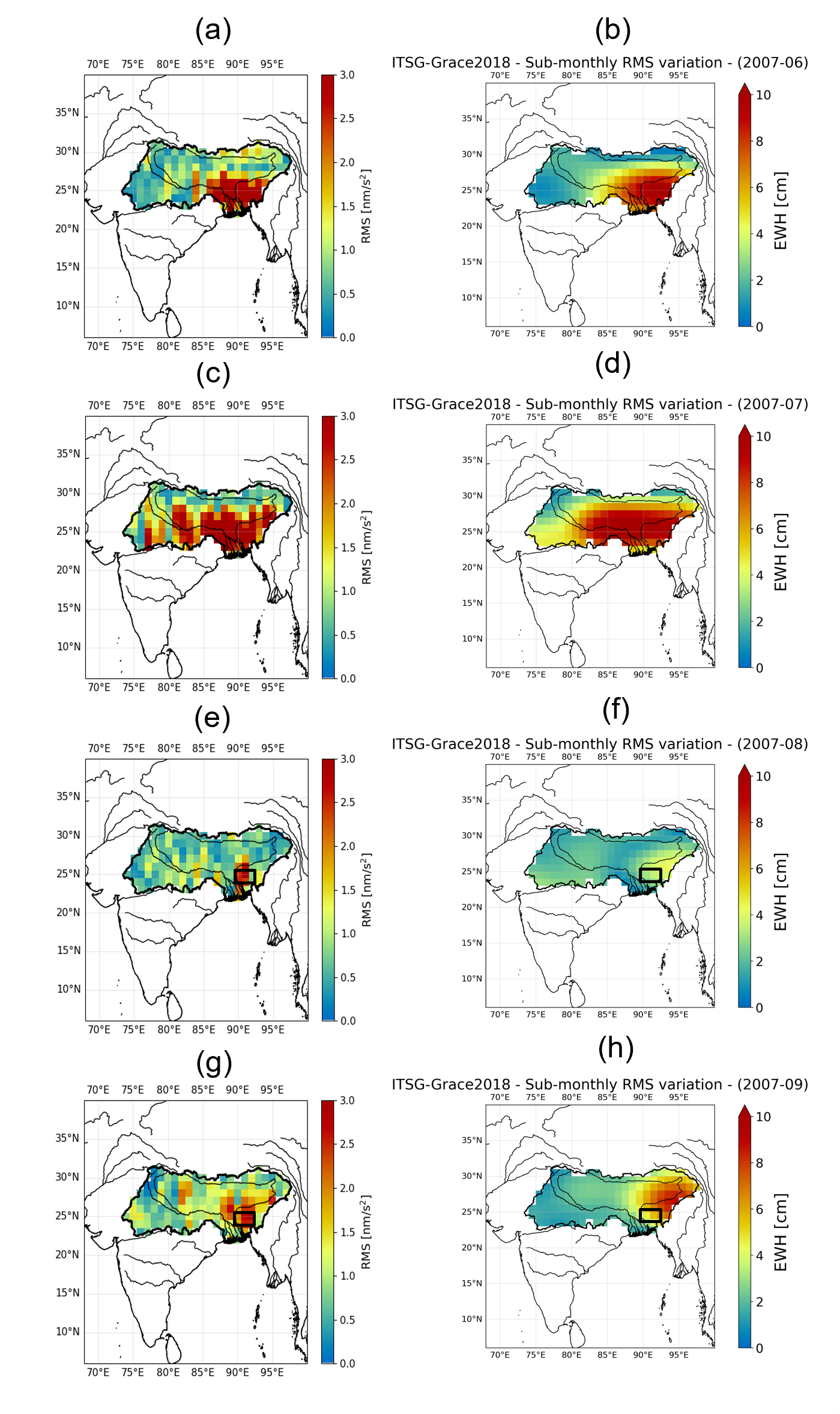}
    \caption{Consecutive monthly grids from June to September 2007 (top to bottom) for the Ganges-Brahmaputra river basin. The left column shows the LGD post-fit residuals RMS (1°x1°), and in the right column the sub-monthly variations in Equivalent Water Heights (EWH) in cm of daily Kalman filtered solutions from ITSG-Grace2018.}
    \label{fig:spatial_maps_GB}
\end{figure}

As performed in the previous section, we compare LGD post-fit residuals to the sub-monthly RMS variations (Equation \ref{eq:RMS_var}) in the ITSG-Grace2018 daily solutions as can be seen in Figure \ref{fig:spatial_maps_GB}. We confirm the findings from previous research \cite{Gouweleeuw2018}, which indicated that these daily Kalman-filtered solutions exhibit sub-monthly variability due to the abrupt flooding events from June to September, as shown in Figure \ref{fig:spatial_maps_GB} (left column). Notably, for June and July (Figures \ref{fig:spatial_maps_GB}a,b and \ref{fig:spatial_maps_GB}c,d), sub-monthly variabilities reach values of 10 cm in EWH coinciding both with LGD variations and the known regions of significant rainfall from 500 to even 1000 mm (refer to Figure 4 by \citet{Islam2010}). The opposite is true for August (Figures \ref{fig:spatial_maps_GB}e and \ref{fig:spatial_maps_GB}f), where both the rainfall and these variabilities reach a minimum, suggesting that monthly models adequately represent the overall water storage fluctuations in the region, as also evidenced by the lower observed RMS in the post-fit residuals.

Furthermore, we observe significant spatial correlations with Pearson correlation coefficients of 0.81 and 0.78 between the LGD post-fit residuals and the sub-monthly RMS variabilities in the daily solutions for June and July, whereas, for August and September, these are 0.38 and 0.56, respectively. 
As performed in Section \ref{subsec:cyclones}, to make this analysis more comparable between both variables, we apply a 2D Gaussian filter (with a radius of 6 deg and $\sigma=2$ to match the smoothing applied on the gridded daily solutions) to the post-fit residuals grid. Using this method, the spatial correlations increase to 0.92 and 0.95 for June and July, and 0.62 and 0.82 for August and September, confirming a better match with the daily solutions' spatial resolution. We refer the reader to Figure S5 in the SI to show the effect of the standard deviation parameter $\sigma$ on the smoothing process. By utilising the L1B 5s sampling rate, we seem to better localise the two flood peaks in August and September in the smaller Meghna River, centred at (24°N, 91°E) \cite{Islam2010}, highlighted in Figures \ref{fig:spatial_maps_GB}e and \ref{fig:spatial_maps_GB}f, whereas the daily solutions do not and are known to have a limited spatial resolution \cite{Gouweleeuw2018}. This is further confirmed by analysing the instantaneous or along-orbit post-fit residuals, which show localised variations near the flooded region in both the ascending and descending tracks. Finally, a longitudinal phase shift of 1-4° is evident between the spatial variations in post-fit residuals and the daily solutions, which we speculate is attributed to the incorporated hydrological model statistics in the Kalman filtering constraints and pre-applied spatial smoothing.

\begin{figure}[htp]
    \centering
    \includegraphics[width=1\textwidth]{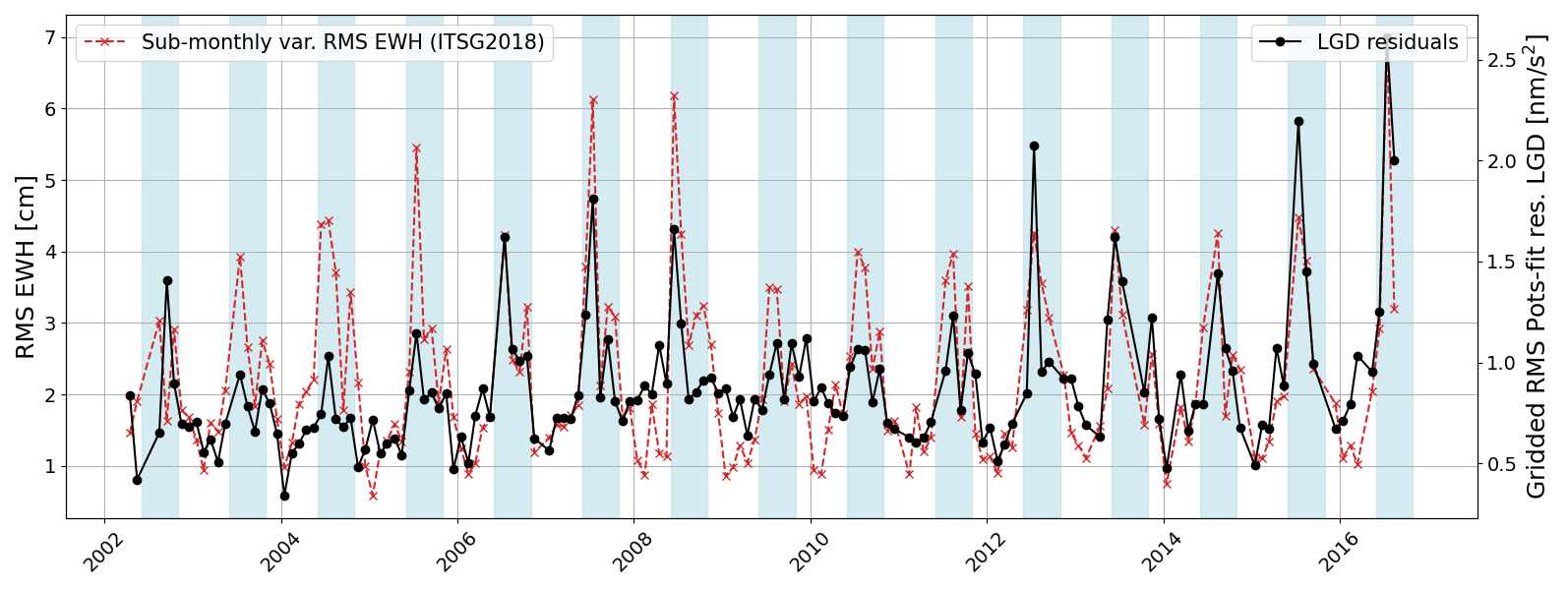}
    \caption{Time series of gridded RMS LGD post-fit residuals, and RMS variability w.r.t. monthly mean of ITSG daily solutions in the Ganges-Brahmaputra basin from April 2002 to August 2016. Shaded regions depict monsoon seasons from June to mid-October.}
    \label{fig:rms_time_func_itsg}
\end{figure}

\begin{figure}[htpb]
    \centering    
    \includegraphics[width=0.625\textwidth]{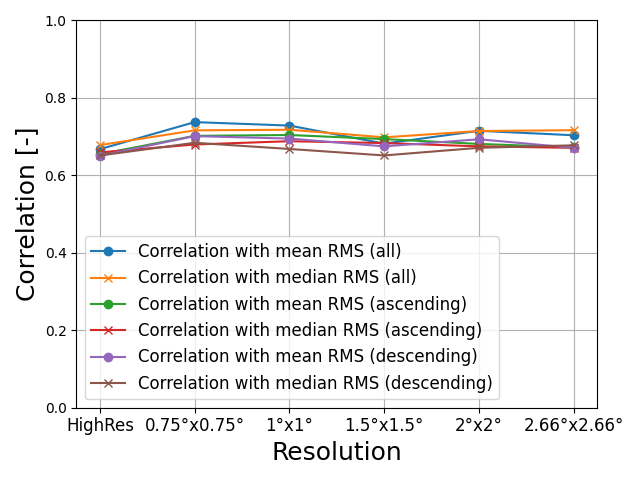}
    \caption{Sensitivity analysis of the correlation coefficient between the RMS of post-fit residuals and the sub-monthly variability in ITSG daily solutions, as influenced by spatial resolution, ground-track type, and statistical estimator. The term 'HighRes' denotes the case in which the post-fit residuals were not spatially gridded.}
    \label{fig:sens_analysis}
\end{figure}

After examining these single flooding events, we conduct the same aggregate statistical analysis described in Section \ref{subsec:oceanic}. This analysis is applied to both the gridded LGD residuals and the daily solutions' variability, resulting in Figure \ref{fig:rms_time_func_itsg}.
Two key observations can be made from Figure \ref{fig:rms_time_func_itsg}. First, there is significant sub-monthly variability in both post-fit residuals and daily solutions during the monsoon seasons, which are indicated by the shaded areas. Second, a high correlation of 0.73 exists between these two time-series, confirming that post-fit residuals effectively reflect sub-monthly variabilities and that these are a consistent seasonal phenomenon that must be accounted for in regional mass change models. Finally, we performed a sensitivity analysis to assess the robustness of our results. This involved varying the spatial resolution of the gridded LGDs, separating the ground tracks into ascending and descending categories, and using either the mean or median as statistical estimators, as illustrated in Figure \ref{fig:sens_analysis}.

In Figure \ref{fig:sens_analysis}, we found that changes in spatial resolution resulted in relatively insignificant alterations in correlation. However, separating the data by ground-track type and using different statistical estimators led to a correlation change ranging from 0.05 to 0.09. Additionally, we truncated the GRACE mission data from July 2003 to eliminate potential issues related to the one SCA phase. This adjustment yielded a slight increase in correlation coefficients of approx. 0.025. Overall, these changes remain relatively minor, showing the robustness of the analysis.

\begin{figure}[htp]
    \centering
    \includegraphics[width=1.\textwidth]{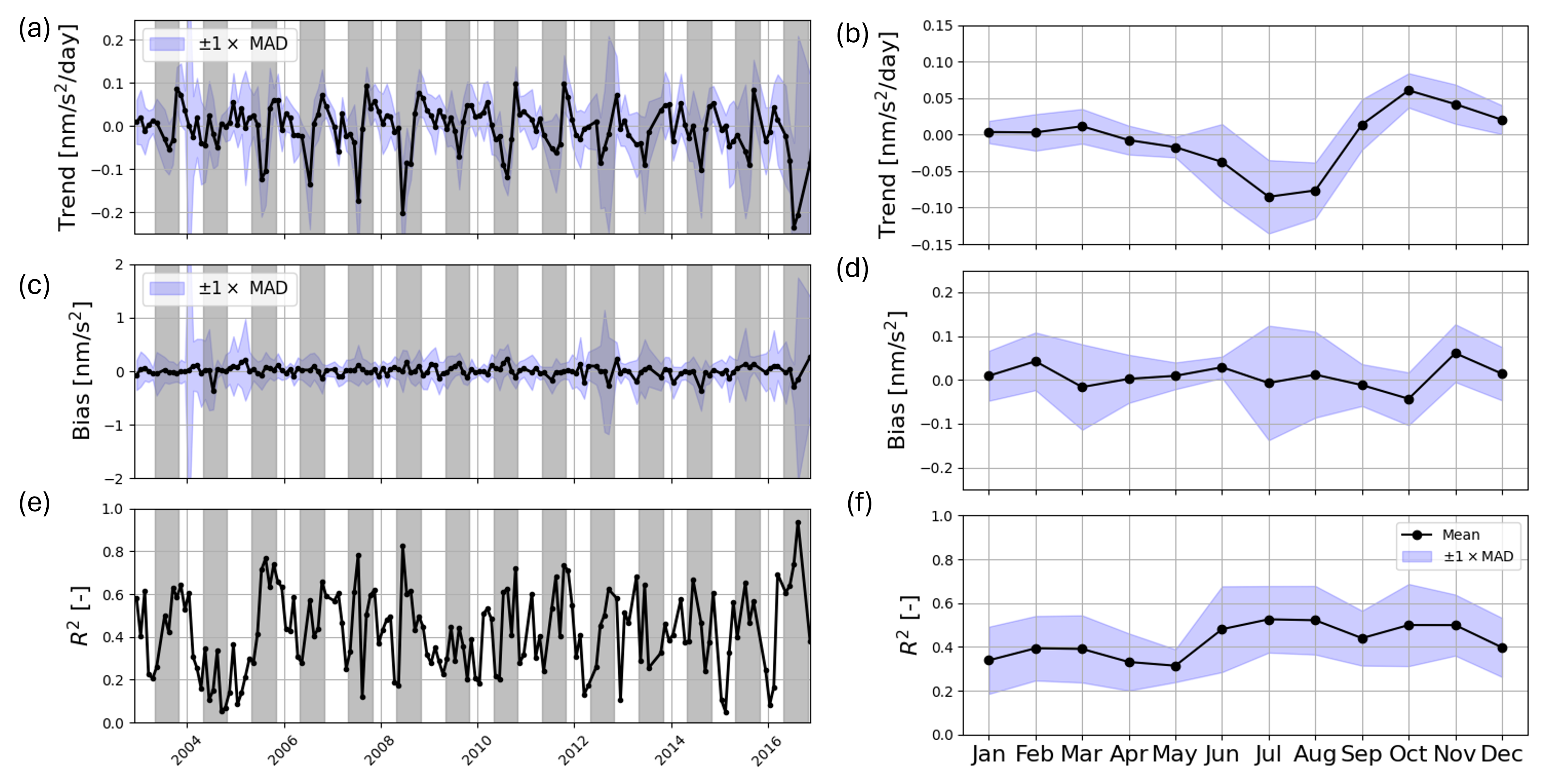}
    \caption{The left column shows the time series of gridded linear regression variables such as trend (\textbf{a}), bias (\textbf{c}) and $R^2$ (\textbf{e}) related to LGD post-fit residuals for the entire Ganges-Brahmaputra river basin. Shaded regions depict monsoon seasons from June to mid-October. The right column shows the average characteristic per month for the trend (\textbf{b}), bias (\textbf{d}) and $R^2$ (\textbf{f}).}
    \label{fig:time_series_lin_reg_GBM}
\end{figure}

Finally, we present the results of the complete time series of sub-monthly LGD linear regression characteristics (trend, bias, and $R^2$) using the methodology presented in Section \ref{sec:methods} in Figure \ref{fig:time_series_lin_reg_GBM} to characterize the periodic nature of the sub-monthly signals.

From Figure \ref{fig:time_series_lin_reg_GBM}, we can observe a seasonal pattern in the trend signal (Figure \ref{fig:time_series_lin_reg_GBM}a). This pattern is characterized by a sharp transition from negative values during June to August, to positive values in September and October (Figure \ref{fig:time_series_lin_reg_GBM}b). The monsoon flood events of 2007 exhibit one of the most significant trend changes, alongside those in 2006, 2008, and 2016, with a shift from -0.17 to 0.1 nm/s²/day. This change corresponds with the increased hydrological activity (i.e. rainfall and flooding) noted during these events \cite{Islam2010}. Additionally, the adjusted \( R^2 \) values show substantial peaks, reaching between 0.6 and 0.95 (Figure \ref{fig:time_series_lin_reg_GBM}e), with an average monthly maximum of 0.56 observed during the monsoon seasons (Figure \ref{fig:time_series_lin_reg_GBM}f). Lastly, as discussed in Section \ref{subsec:temp}, the mid-month bias does not display any discernible temporal pattern (with a slight increase in MAD spread during July), as evidenced by Figures \ref{fig:time_series_lin_reg_GBM}c and \ref{fig:time_series_lin_reg_GBM}d. These results therefore affirm that the L1B post-fit residuals indeed capture significant seasonal signals associated with sub-monthly land hydrology which can be well described by a spatial-temporal linear regression.

\subsection{Global temporal correlations between post-fit residuals and ITSG daily solutions}
\begin{figure}[ht]
    \centering
    \includegraphics[width=0.875\textwidth]{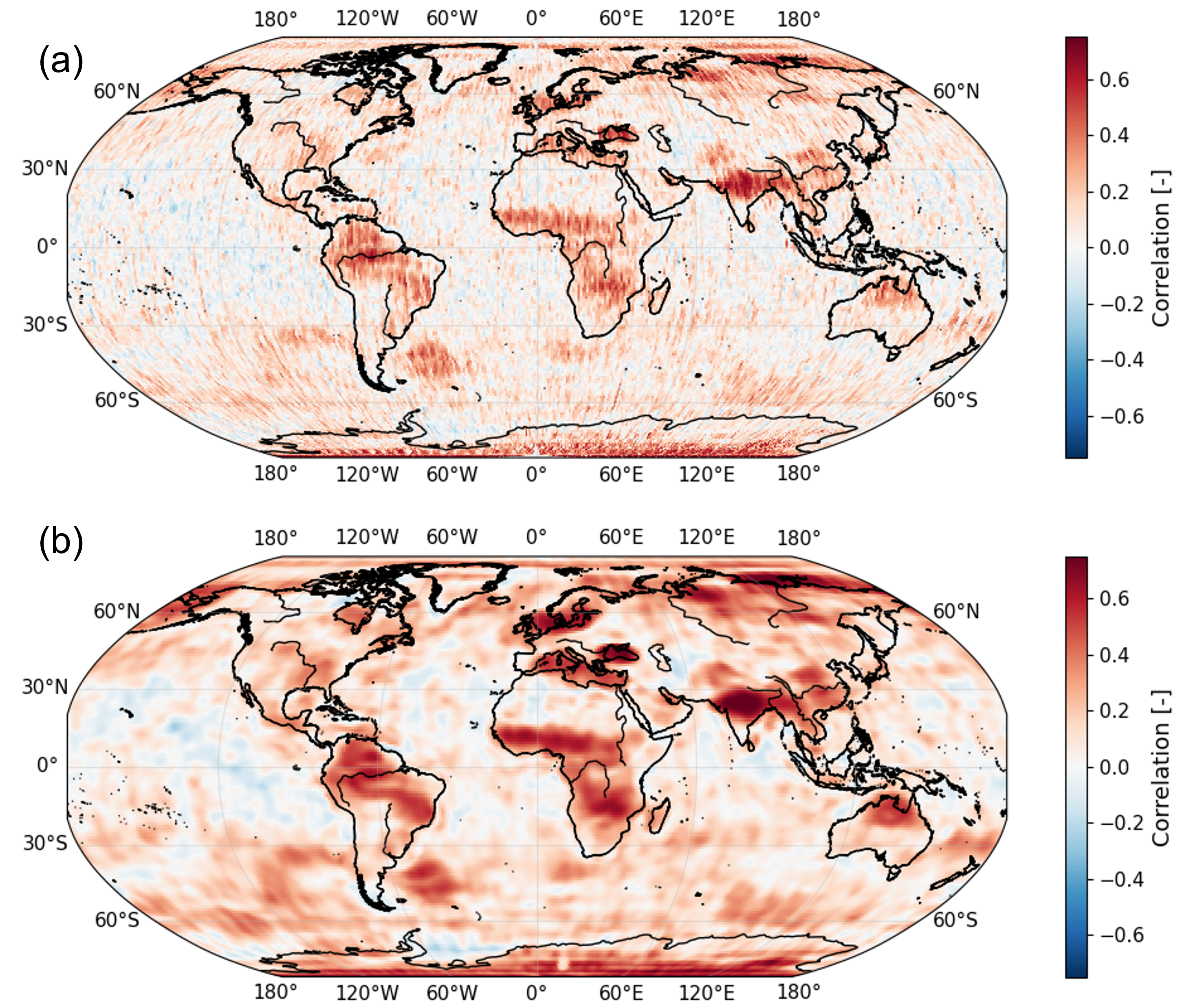}
    \caption{Global overview of the spatial distribution of temporal correlations between post-fit residuals as Line-of-Sight Gravity Differences (1°x1°) and RMS sub-monthly variability of ITSG-Grace2018 daily solutions (from July 2003 to August 2016). (\textbf{a}) and (\textbf{b}) depict respectively the temporal correlations without and with Gaussian spatial smoothing (radius of 6 deg and $\sigma=2$).}
    \label{fig:global_corrs}
\end{figure}

Having presented the results of individual sub-monthly geophysical categories, we will now provide a global overview of the spatial distribution of the temporal correlations between the monthly gridded post-fit residuals and the sub-monthly RMS variability of daily solutions, as illustrated in Figure \ref{fig:global_corrs}.

As anticipated, applying Gaussian spatial smoothing increases the temporal correlations both locally and globally (as seen in Figure \ref{fig:global_corrs} from part a to part b). This improvement occurs as we better align with the original spatial resolution of the daily solutions, which is approximately 500 km. This resolution was effectively reduced after implementing spatial smoothing. 

We observe significant correlations (above 0.5) in several major river basins. In South America, the Amazon, Orinoco, and La Plata river basins showed correlations between 0.5 and 0.6. In Central Africa, notable correlations (up to 0.65) were found in the Chad, Nile, and Congo river basins. In Asia, the Ganges-Brahmaputra and Indus River basins also demonstrated strong correlations. In Northern Australia, we identified correlations reaching up to 0.6 related to the aforementioned hydro-meteorological sub-monthly variations due to cyclonic activity. Of all river basins, the Indus and Ganges-Brahmaputra basins exhibited the highest temporal correlations, with values ranging from 0.60 to as high as 0.75.

Temporal correlations are not limited to land hydrology sub-monthly signals; they can also be seen in the Argentine ocean basin, the Mediterranean, and the North Sea, as well as along the western coast of Canada and Alaska, and the northern coast of Russia. The most noticeable oceanic correlations are within the Argentine basin which are related to the high-frequency component of the rotating dipole mode around the Zapiola rise.  Sub-monthly glaciology signals predominantly appear in Antarctica, with a limited presence in Greenland (with correlations capped at 0.4). 

Lastly, we can identify several regions with negative or no temporal correlations, primarily in the oceans, which we speculate may be related to remaining KBR noise in the post-fit residuals that dominate compared to the geophysical signal. Given the high correlations between post-fit residuals and daily solutions in the highly active hydrological river basins, this approach further validates the assumed geophysical constraints implemented within the ITSG-Grace2018 daily models.

\section{Conclusions}\label{sec:conclusions}
In this study, we examined the use of K-band range-rate post-fit residuals from the GRACE mission provided by CSR, which were transformed into  Line-of-Sight Gravity Differences (LGD), to identify and characterise high-frequency mass change signals. The purpose of this is to answer: 

\textit{How can high-frequency geophysical signals within GRACE Level-1B post-fit residuals be identified, characterised, and validated using spatial analysis and daily solutions?}

Our findings confirmed previous observations related to the geophysical frequency range of [0.9, 11] mHz of sub-monthly signals in post-fit residuals while significantly advancing the understanding of these residuals by providing detailed geophysical attribution using a spatial-temporal linear regression and Root-Mean Squared characterisation and validation of their sub-monthly nature.

We focused on three main categories of high-frequency signals: oceanic, meteorological, and hydrological processes. In the Zapiola region, we detected significant high-frequency oceanic mass change signals potentially linked to the dynamics of a rotating dipole-like mode. However, a comparison of Sea Level Anomalies and LGD variations through statistical analysis showed no strong correlations (below 0.25), suggesting that further research is needed. 

During Cyclone Oswald, we identified significant meteo-hydrological signals in the Gulf of Carpentaria (GoC), supporting earlier research that linked LGD anomalies to rapid pressure changes in this region. Our extended analysis further revealed that the areas of LGD anomalies aligned closely with the cyclone’s path and associated precipitation patterns. For the first time, we validated the sub-monthly nature of these anomalies using RMS variations from ITSG-Grace2018 daily solutions. The spatial correlations between the LGD anomalies and these daily solutions reinforced the interpretation of sub-monthly variability. Additionally, we developed a preliminary proto-mass change model that approximates the cyclone as a point mass, which effectively localised two cyclonic sub-monthly signals during one of the cyclone’s peak intensities over the GoC. However, limitations in this simplified model became evident as the majority of cases have partial to no alignment with post-fit residuals, highlighting the necessity of a more advanced mass change modelling framework.

Lastly, we also analysed LGD residuals in the Ganges-Brahmaputra river basin during the 2007 monsoon season and we showed that regions of significant LGD residuals variations correlated with sub-monthly variations in the ITSG-Grace2018 daily solutions. We extended this analysis from July 2003 to August 2016 and demonstrated first that there is a significant correlation of 0.67-0.74 between these two time-series and secondly, that sub-monthly variabilities both in LGD residuals and daily solutions tend to peak during monsoon seasons. Additionally, the 5-second sampling rate of the Level-1B data enabled us to potentially better localise critical flooding peaks in the Meghna river, which was not achievable with the coarse spatial resolution of the daily solutions. 

This study provides the first extensive geophysical characterisation of high-frequency signals within GRACE post-fit residuals—specifically, those from the widely used CSR RL06 Level-2 data product. Previous work primarily suggested that post-fit residuals contained sub-monthly signals but lacked systematic attribution to specific geophysical processes or robust validation methods. Here, we introduced spatial plots of statistics and linear regression parameters, and we performed a cross-validation using the ISTG-Grace2018 daily solutions for the first time to demonstrate their sub-monthly nature. This highlights the broader potential of these residuals to offer improved temporal and spatial resolution for studying rapid changes in Earth's gravity field.

Overall, our findings underscore the value of GRACE post-fit residuals as an intermediate product, with the potential to enhance the temporal and spatial resolution of gravity-based mass change monitoring. Future research will focus on leveraging these residuals to develop a framework for high-frequency mass change modelling.

\bibliographystyle{unsrtnat}

\end{document}